\documentclass[fleqn,usenatbib]{mnras}

\usepackage{newtxtext,newtxmath, comment}

\usepackage[T1]{fontenc}

\DeclareRobustCommand{\VAN}[3]{#2}
\let\VANthebibliography\thebibliography
\def\thebibliography{\DeclareRobustCommand{\VAN}[3]{##3}\VANthebibliography}

\usepackage{graphicx}	
\usepackage{amsmath}	
\usepackage{footnote}

\title[The VIRA Retrieval Framework]{VIRA: An Exoplanet Atmospheric Retrieval Framework for JWST Transmission Spectroscopy}

\author[Constantinou \& Madhusudhan]{
Savvas Constantinou,$^{1}$\thanks{E-mail: sc938@ast.cam.ac.uk}
Nikku Madhusudhan$^{1}$\thanks{E-mail: nmadhu@ast.cam.ac.uk}
\\
Institute of Astronomy, University of Cambridge, Madingley Road, Cambridge CB3 0HA, UK\\
}

\date{Accepted 21$^{\mathrm{st}}$ February 2024. Received 20$^{\mathrm{th}}$ February 2024; in original form 3$^{\mathrm{rd}}$ August 2023}

\pubyear{2024}

\begin{document}
\label{firstpage}
\pagerange{\pageref{firstpage}--\pageref{lastpage}}
\maketitle

\begin{abstract}

JWST observations are leading to important new insights into exoplanetary atmospheres through transmission spectroscopy. In order to harness the full potential of the broad spectral range and high sensitivity of JWST, atmospheric retrievals of exoplanets require a high level of robustness and accuracy in the underlying models. We present the VIRA retrieval framework  which implements a range of modelling and inference capabilities motivated by early JWST observations of exoplanet transmission spectra. This includes three complementary approaches to modelling atmospheric composition, three atmospheric aerosol models, including a physically-motivated Mie scattering approach, and consideration of correlated noise. VIRA enables a cascading retrieval architecture involving a sequence of retrievals with increasing  sophistication. We demonstrate VIRA using a JWST transmission spectrum of the hot Saturn WASP-39~b in the $\sim$1-5~$\mu$m range. In addition to confirming prior chemical inferences, we retrieve molecular abundances for H$_2$O, CO, CO$_2$, SO$_2$ and H$_2$S, resulting in super-solar elemental abundances of log(O/H)=$-2.0\pm0.2$, log(C/H)=$-2.1\pm0.2$ and log(S/H)=$-3.6\pm0.2$, along with C/O and S/O ratios of $0.83^{+0.05}_{-0.07}$ and $0.029^{+0.012}_{-0.009}$, respectively, in the free chemistry case. The  abundances correspond to $20.1^{+10.5}_{-8.1}\times$, $28.2^{+16.3}_{-12.1}\times$ and $20.8^{+10.3}_{-7.5}\times$ solar values for O/H, C/H and S/H, respectively, compared to C/H $= 8.67\pm0.35 \times$ solar for Saturn. Our results demonstrate how JWST transmission spectroscopy combined with retrieval frameworks like VIRA can measure multi-elemental abundances for giant exoplanets and enable comparative characterisation with solar system planets. 

\end{abstract}




\section{Introduction}
\label{sec:intro}

A year into the JWST era, the revolution in our understanding of exoplanetary atmospheres is well underway. Early transmission spectroscopy observations, such as those of WASP-39~b, have probed novel spectral regions and led to detections of numerous chemical species, such as CO$_2$, H$_2$O, SO$_2$, H$_2$S and CO \citep{Ahrer2023, Alderson2023, Constantinou2023, Feinstein2023, Rustamkulov2023, Niraula2023}. Detecting and measuring the abundances of such a broad range of chemical species can enable new insights into atmospheric processes, as demonstrated with CO$_2$ and SO$_2$ in the hot Saturn WASP-39~b \citep{Tsai2023} and CO$_2$ and CH$_4$ in the candidate Hycean world K2-18~b \citep{Madhusudhan2023}.

JWST comes after more than two decades of pioneering work enabled by the Hubble Space Telescope (HST). The beginning of the HST era was marked with the first detections of exoplanetary atmospheres and chemical species in them \citep{Charbonneau2002, VidalMadjar2003, VidalMadjar2004}.  By the time JWST was launched, HST observations routinely led to robust detections of H$_2$O in the infrared for a wide range of planets as well as Na and K in the optical for irradiated gas giants \citep{Madhusudhan2019}. 

Beyond chemical detections, HST-era observations have revealed the significant impact atmospheric aerosols can have on a planet's transmission spectrum. HST and ground-based observations in the optical for numerous planets have found pronounced slopes towards shorter wavelengths arising from Mie-scattering atmospheric aerosols, which can fully or partially obscure spectral features from Na and K \citep[e.g.][]{Pont2008, NIkolov2015, Spyratos2023}. Meanwhile infrared HST observations have found significantly muted H$_2$O features, which has been attributed to relative H$_2$O depletions and/or masking due to cloud absorption \citep[e.g.][]{Pont2013, Deming2013, Kreidberg2014a, Knutson2014, Madhusudhan2014b, Sing2016}.

State of the art atmospheric retrievals are a central part of remote sensing with JWST, having already played a key role in the HST era \citep[e.g.][]{Madhusudhan2009, Madhu2011_w12b, Benneke2012, Line2013, Madhusudhan2014b,  Fischer2016, Macdonald2017, Barstow2017, Tsiaras2018, Gandhi2018, Pinhas2019, Madhusudhan2019, Welbanks2019b, Molliere2019, Changeat2022, Edwards2023}. By thoroughly exploring the available parameter space, Bayesian retrievals enable robust constraints for a wide variety of atmospheric properties, such as the abundances of different chemical species, the properties of atmospheric aerosols and the atmospheric temperature structure. Modern atmospheric retrievals can also be used to assess the confidence with which chemical species and other atmospheric properties are detected through Bayesian model evidence comparisons \citep{Benneke2013}.

The HST era of transmission spectroscopy has also led to several surprising findings, such as an inference of general H$_2$O depletion for several exoplanets relative to the solar system metallicity trend \citep{Madhusudhan2014a, Pinhas2019, Barstow2017, Welbanks2019b}. Moreover, this depletion persists even when accounting for the expected increase in overall atmospheric metallicity with lower planet masses and the presence of clouds. At the same time, observed Na and K abundances did not show such a depletion, instead indicating a mass-metallicity trend that is consistent with that obtained for solar system planets using their CH$_4$ abundances \citep{Welbanks2019b}. On the other hand, various studies have considered different combinations of atmospheric spectra, e.g. transmission spectra with HST/WFC3 alone and/or emission spectra with HST or Spitzer, finding a diversity of atmospheric inferences \citep{Fischer2016, Tsiaras2018, Changeat2022, Edwards2023}.

JWST constitutes a generational leap in observational capabilities over HST, with a collecting area over 6 times larger, as well as a substantially larger spectral range (0.6-28~$\mu$m) and resolution (up to R$\sim$3000) \citep{Beichman2014, Stevenson2016b, Batalha2017a, Kalirai2018}. This has already been demonstrated for irradiated gas giant planets such as WASP-39~b as mentioned above, as well as WASP-96~b \citep{Radica2023}. The JWST transmission spectra of both planets contain several molecular absorption features leading to highly confident detections of H$_2$O, as well as multiple other molecules in the case of WASP-39~b.

Early JWST observations of irradiated gas giants have also indicated the continuing role atmospheric aerosols are likely to play in the JWST era \citep{Constantinou2023, Feinstein2023}. Thanks to the significantly larger spectral coverage of JWST, it is possible to probe how the opacity contributions of aerosols vary with wavelength, providing constraints on their composition, modal particle size and other physical properties \citep{Wakeford2015, Pinhas2017}. This means that the existing practice of modelling aerosols as a grey opacity, which was appropriate for the narrower wavelength range of HST, may no longer be sufficient in some cases. Given that extinction due to aerosols can mask molecular spectral features to different extents, models that do not accurately capture the wavelength dependence of this truncation may also obtain inaccurate abundance estimates. On the other hand, atmospheric retrievals capable of modelling Mie scattering aerosols are able to constrain their composition and physical properties, with sufficiently precise observations \citep{Mai2019, Ormel2019, Lacy2020a, Constantinou2023}.

\begin{figure*}
    \centering
    \includegraphics[angle=0,width=0.85\textwidth]{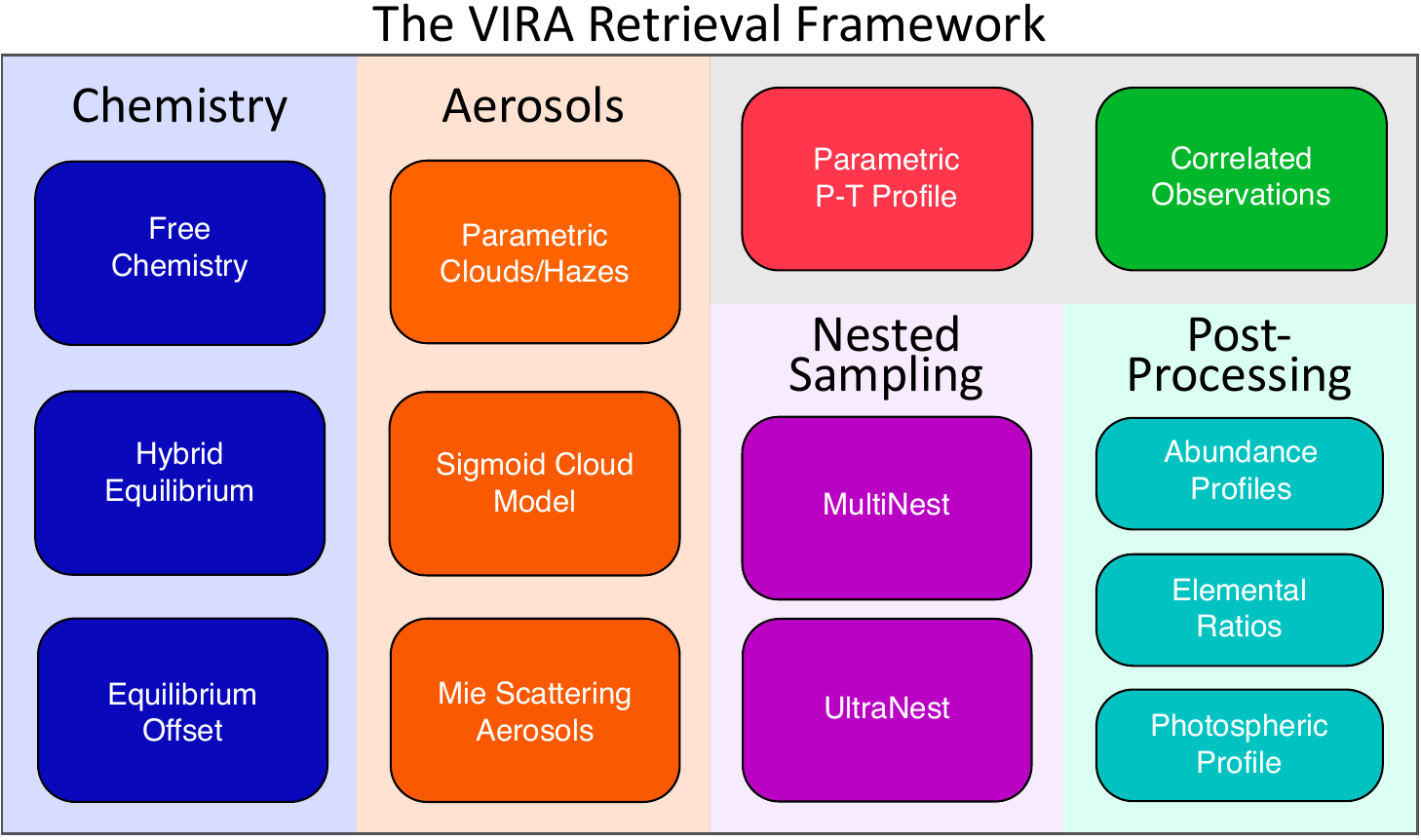}
    \caption[A schematic representation of the modelling and statistical techniques implemented in the VIRA retrieval framework.]{A schematic representation of the modelling and statistical techniques available to the VIRA retrieval framework. The framework's features are intended to be used as part of a cascade of multiple retrievals, with the results of prior retrievals being used to inform subsequent more refined runs.}
    \label{vira_fig:VIRA_diagram}
\end{figure*}

Transmission spectroscopy with JWST also enables simultaneous detections of multiple chemical species, enabling unprecedented insights into the physical and chemical processes driving them. This was first demonstrated with the detection of SO$_2$ in the atmosphere of WASP-39~b \citep{Alderson2023, Rustamkulov2023}, which was found to be the product of photochemical processes \citep{Tsai2023}. Photochemistry and other equilibrium and disequilibrium processes can affect the relative abundances of different chemical species. In order to infer the presence of such processes, atmospheric retrievals must have significant flexibility, encompassing a broad variety of atmospheric compositions, while robustly assessing how such compositions deviate from equilibrium expectations.

Given the high precision of JWST observations, particularly of irradiated gas giants possessing extended atmospheres, the accuracy of traditional retrievals may be at risk. As such, atmospheric models used in retrievals must reach a high level of sophistication to avoid inaccurate results. This includes considering atmospheric composition that is non-uniform, which has been shown to be important \citep{Rocchetto2016}. Atmospheric retrievals must be able to consider both molecules whose abundances vary as a result of equilibrium chemistry, such as CO and CH$_4$, as well as species that are significantly affected by disequilibrium processes, such as SO$_2$ mentioned above. Retrievals must therefore be able to consider non-uniform vertical mixing ratio profiles, without solely relying on the assumption of thermochemical equilibrium.

In order to make the most of the high density of information encoded in JWST spectra, the atmospheric models used in retrievals must strike a balance between model flexibility and sophistication. The choice of model must be motivated both by the precision and wavelength coverage of the data, as well as what is physically plausible for each planet.  A systematic approach to managing modelling complexity depending on the observations is therefore necessary, in order for the retrieval to be computationally tractable, avoid unphysical constraints and to make the most of the available data.

The high precision and resolution of JWST observations mean that it is possible to resolve highly detailed spectral features, which are highly informative to atmospheric retrievals. In this regime any correlations between datapoints can become important and affect the accuracy of the retrieved atmospheric properties. This is particularly important for the Near Infrared Imager and Slitless Spectrograph (NIRISS), which has been shown to produce observations with significant correlations between datapoints that are widely separated in wavelength \citep{Holmberg2023}.

High-quality JWST observations can enable extremely precise constraints for atmospheric properties including composition, temperature structure and the properties of aerosols. Such precise constraints must be interpreted in the context of the specific atmospheric region. The exact altitude probed by transmission spectra, however, depends on the abundance of gaseous species, the presence of clouds and hazes, and also the particular spectral region being probed \citep{Madhusudhan2019}. As such, it is important that atmospheric retrievals can establish where this effective photosphere lies on a case-by-case basis.

Motivated by the paradigm-shifting potential of JWST, we present VIRA, an atmospheric modelling and retrieval framework expanding on the legacy of AURA-family retrieval codes. Seeking to make maximal usage of JWST observations, VIRA follows a new paradigm in retrieval methodology, whereby a cascade of retrievals are carried out, progressively exchanging generality for modelling sophistication. In this process, VIRA implements an array of atmospheric models and parametrisations which are used over the course of the retrieval cascade. This includes three parametrisations for atmospheric composition, including a disequilibrium based approach to modelling atmospheric abundances and their vertical distributions, as well as three approaches to modelling atmospheric aerosols, of which two are parametric in nature and one relies on a physical Mie scattering model. VIRA also implements a robust consideration of correlated noise in observations, which was found to be significant particularly for NIRISS data \citep{Holmberg2023}. Lastly, in light of the complex parameter space that atmospheric models can present, VIRA can utilise two nested sampling algorithms to ensure the robustness of the retrieved constraints.

This work presents the atmospheric modelling and statistical capabilities of the VIRA retrieval framework in Section \ref{vira_sec:VIRA_framework}, focusing on the new features implemented over prior AURA-family frameworks. In particular, we present the three complementary atmospheric models in Section \ref{vira_subsec:atmospheric_composition}, while in Section \ref{vira_subsec:nongrey_clouds} we present the three available atmospheric aerosol models. We subsequently describe the statistical framework implemented to consider correlated noise in observations in Section  \ref{vira_subsec:niriss_covariance}, summarise the parametric pressure-temperature (P-T) profile used in section \ref{vira_subsec:PT_profile} and lastly present the two nested sampling algorithms VIRA uses in Section \ref{vira_subsec:ultranest}. We then demonstrate the capabilities of VIRA by carrying out a thorough retrieval analysis of the flagship Cycle 1 Early Release Science observations of the hot Saturn WASP-39~b in Section \ref{vira_sec:WASP-39b_retrievals}. After an overview of prior atmospheric constraints obtained with JWST observation in section \ref{vira_subsection:prior_w39b_inference}, we present each step of the retrieval cascade in detail, starting with free chemistry retrievals in Section \ref{vira_subsec:free_chemistry_retrievals}, hybrid equilibrium retrievals in Section \ref{vira_subsec:hybrid_equilibrium} and lastly equilibrium offset retrievals in Section \ref{vira_subsec:equilibrium_offset_rets}. We finally discuss and summarise the VIRA retrieval framework the retrieved atmospheric properties for WASP-39~b in Section \ref{vira_sec:discussion}.

\section{The VIRA Retrieval Framework}
\label{vira_sec:VIRA_framework}

The VIRA retrieval framework consists of a forward model generator coupled with a Bayesian parameter estimator, similarly to most state-of-the-art retrieval frameworks \citep{Madhusudhan2019}. The VIRA forward model generator is based on that used in AURA \citep{Pinhas2018}, which has been subsequently built upon to include non-H$_2$ dominated atmospheres in Aurora \citep{Welbanks2021} and a 3-dimensional treatment of the atmosphere in AURA-3D \citep{Nixon2022}. VIRA builds upon the functionality of prior AURA-family retrieval frameworks. The forward model generates transmission spectra by computing line-by-line radiative transfer for a 1D, plane-parallel and hydrogen-dominated atmosphere under hydrostatic equilibrium. The cross-sections of gaseous species are computed following \citet{Gandhi2017} and \citet{Gandhi2020}, at 0.1~$cm^{-1}$ resolution. The details of the opacity sources are described in Section \ref{vira_subsubsec:free_sigmoid_ret}. In this section, we present the new features that VIRA offers in order to rise to the challenges of the JWST era, which are summarised in Figure \ref{vira_fig:VIRA_diagram}. These include several advancements in the physical accuracy of the retrieved atmospheric model as well as more sophisticated approaches in handling observations and sampling the available parameter space.

\subsection{A Cascading Retrieval Architecture}
\label{vira_subsec:cascading_architecture}

\begin{figure*}
    \centering
    \includegraphics[angle=0,width=1.0\linewidth]{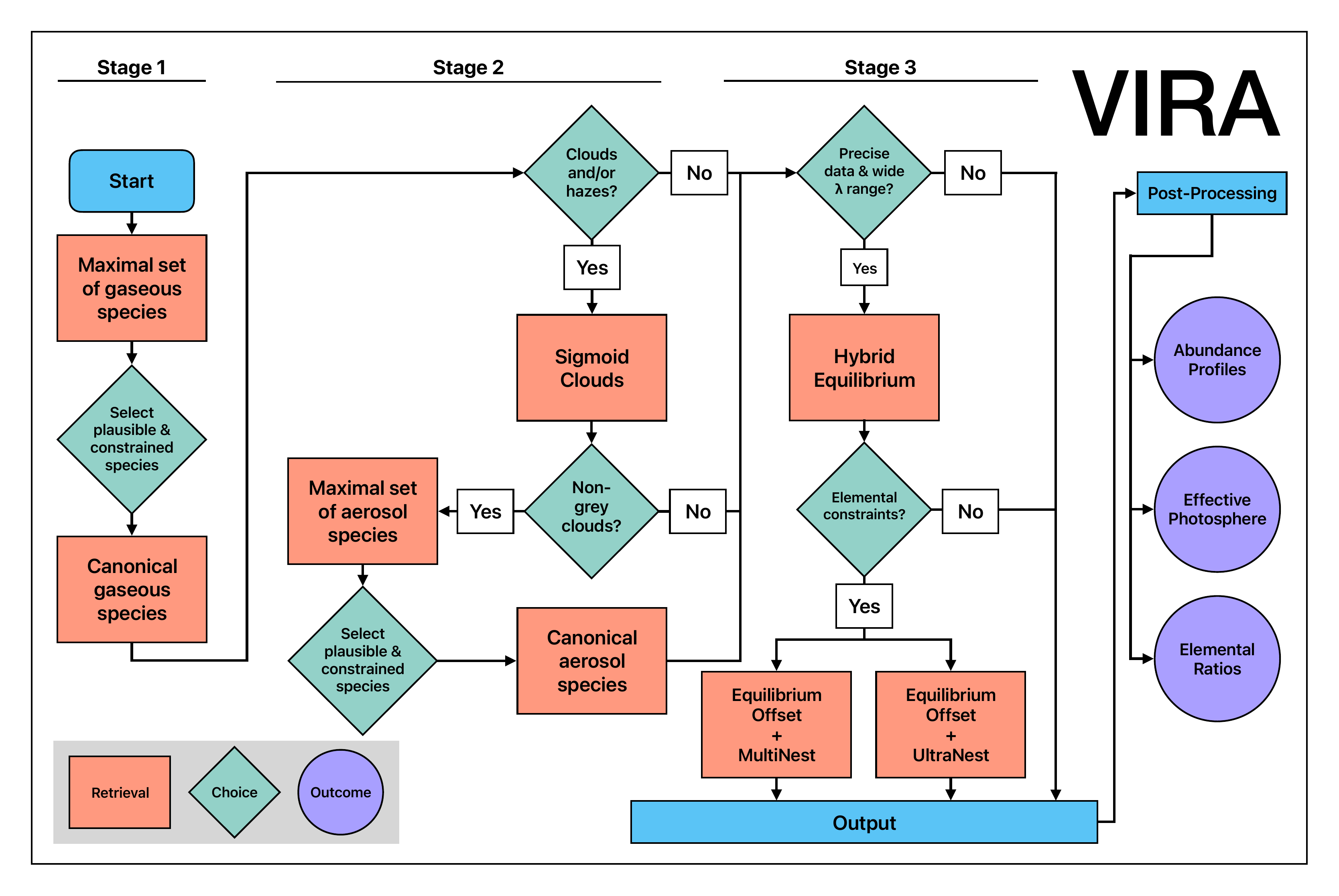}
    \caption{ Flowchart showing how a VIRA retrieval cascade may proceed. Stage 1 establishes detectable gaseous species and is equivalent to AURA retrievals. Stage 2 establishes the nature and properties of atmospheric aerosols, while Stage 3 refines the atmospheric composition constraints.}
    \label{vira_fig:VIRA_flowchart}
\end{figure*}

JWST transmission spectra have already been shown to encode a wealth of information about the observed planet's atmosphere. In order to fully mine all available information and build a complete picture of a planet's atmospheric properties, VIRA is designed as a cascading retrieval architecture. This involves carrying out an array of increasingly sophisticated retrievals, with the results of each step being used to inform the atmospheric models and statistical techniques employed in subsequent retrievals.

This cascading approach applies to numerous aspects of the retrieval framework described in this section. These include the treatment of atmospheric composition discussed in section \ref{vira_subsec:atmospheric_composition}, atmospheric aerosols, presented in section \ref{vira_subsec:nongrey_clouds} and the sophistication of the parameter estimator used, as described in \ref{vira_subsec:ultranest}. Each step in the process balances model generality with accuracy, with the goal of obtaining accurate and robust constraints for the observed planet's atmospheric parameters. The number of steps in this process and the level of sophistication required will ultimately depend on the signal-to-noise ratio and wavelength range of the available observations. 

Figure \ref{vira_fig:VIRA_flowchart} shows how a VIRA retrieval cascade can be carried out. The decisions made at each stage of the cascade are not only motivated by constraints obtained with prior retrievals, but also account for theoretical expectations and physical plausibility. In Stage 1 of the retrieval cascade, we seek to establish which gaseous species should be in the canonical set used in subsequent retrievals. The atmospheric model used involves the free chemistry approach and aerosols modelled as a parametric grey cloud deck and modified Rayleigh-like hazes. This is described in Sections \ref{vira_subsubsec:free_chemistry} and \ref{vira_subsubsec:parametric_cloudsandhazes}, and is equivalent to conventional AURA retrievals. We begin by carrying out a retrieval that includes all available gaseous species. The canonical set of gaseous species is constructed by considering which species have constrained abundances from this first retrieval and/or whose presence is physically plausible. We then carry out a retrieval with only the canonical gaseous species set, which serves as a check on whether the retrieved atmospheric composition constraints are affected by the choice of chemical species present.

Stage 2 of the cascade involves refining the atmospheric aerosol model. If the last retrieval of Stage 1 shows evidence of significant spectral contributions from clouds/hazes, we carry out a retrieval using the more elaborate parametric sigmoid clouds model, described in Section \ref{vira_subsubsec:sigmoid_clouds}. The parametrisation of this model makes it possible to assess if the atmospheric aerosol contributions significantly deviate from grey opacity. If this is indeed the case, we carry out further retrievals using the Mie scattering aerosol model described in \ref{vira_subsubsec:mie_aerosols}. These retrievals are the aerosol equivalent to the selection of gaseous species of Stage 1. This time we seek to establish a canonical set of aerosol species, based on their physical plausibility and the constraints obtained from retrievals.

Stage 3 is the final part of the cascade, and is carried out for transmission spectra that span a significant wavelength range and/or are very precise. The high information content of such datasets may necessitate more sophisticated approaches to atmospheric chemistry, which are explored in this stage. We first carry out a retrieval using the hybrid equilibrium approach, described in section \ref{vira_subsubsec:hybrid_equilibrium_chemistry}. This approach indicates whether or not key elemental abundances such as O/H and C/H can be constrained with the present data and provides first estimates of their values. If such constraints are obtained, they are used as inputs for subsequent retrievals using the equilibrium offset approach described in Section \ref{vira_subsubsec:offset_retrievals}. In order to ensure the robustness of our findings, we carry out retrievals for this final part of the cascade using two nested sampling implementations, discussed in Section \ref{vira_subsec:ultranest}. The results from these retrievals, as well as those obtained from prior runs are then used for the post-processing stage, extracting retrieved vertical abundance profiles, effective photosphere pressures and elemental abundance ratios, as discussed in Section \ref{vira_subsubsec:constraint_postprocessing}.

The present work primarily focuses on more sophisticated approaches to atmospheric chemistry and the spectral effects of atmospheric aerosols. The retrieval cascade however can be further expanded to include additional considerations on a case-by-case basis. Such additions can include stellar heterogeneities from AURA \citep{Pinhas2018}, non-H$_2$ rich atmospheric compositions as used in Aurora \citep{Welbanks2021}, 3D atmospheric effects 
like those in AURA-3D \citep{Nixon2022}, as well as additional P-T parametrisations such as the spline-based approach of \citet{Piette2020c}. In the cascading framework, whether they are included will be determined by evidence provided by simpler retrievals, such as high retrieved chemical abundances or a hot and precisely constraint P-T profile.

\subsection{Atmospheric Composition and Chemistry}
\label{vira_subsec:atmospheric_composition}

Retrieval frameworks in the HST era largely fall in two paradigms, depending on how they model a planet's atmospheric composition \citep[e.g.][]{Madhusudhan2009, Madhusudhan2014b, Macdonald2017, Pinhas2018, Molliere2019, Kitzmann2020, Zhang2020, Welbanks2021, Al-Refaie2021}. On the one hand are so-called ``free'' retrievals, whereby the mixing ratios of each individual chemical species in the model atmosphere is treated as a free parameter. This chemistry-agnostic approach is unaffected by the presence or absence of disequilibrium processes. It can also directly indicate which molecules are detected, while marginalising over the possible mixing ratios of other molecules which are undetected but may be expected to be present. Global compositional parameters such as metallicity and the carbon-to-oxygen (C/O) ratio can be computed by combining the retrieved abundance constraints for detected species. While this approach is highly flexible, the retrieved mixing ratio constraints must be carefully interpreted with regard to their physical plausibility. In its simplest form, such an approach necessitates the assumption of constant-with-altitude chemical mixing ratios, which may no longer be suitable for JWST-quality observations, depending on the specific properties of the observed planet. Previous AURA-family retrieval frameworks all fall under the free retrieval paradigm. Non-constant vertical abundance profiles have been modelled with parametric approaches, necessitating additional free parameters per species \citep[e.g.][]{Changeat2019}, or by offsetting a pre-computed mixing ratio profile per species \citep{Madhusudhan2009}. The top-down approach of free retrievals enables significant flexibility, but does not directly yield constraints on global compositional parameters such as the atmospheric metallicity or elemental abundances.

Equilibrium chemistry retrievals follow the second approach to modelling a planet's atmospheric composition. In this paradigm, retrievals set out to constrain global compositional parameters, such as the atmospheric metallicity, C/O ratio and/or individual elemental abundances. To generate a forward model, a thermochemical equilibrium calculation is performed to obtain individual molecular and atomic mixing ratios. Such an approach does not rely on explicitly including a given set of chemical species, but may not be able to fully explain spectral features that arise due to disequilibrium processes such as photochemistry or vertical mixing. This approach can be further refined by including vertical mixing through a quenching pressure or eddy diffusion parameter \citep[e.g.][]{Molliere2020, Kawashima2021, AlRefaie2022}. Ultimately, the flexibility of such bottom-up approaches is set by the sophistication of the atmospheric model. If a planet's atmosphere is strongly influenced by physical or chemical processes not considered in the model, this may result in inaccurate atmospheric constraints. For instance, frameworks only incorporating vertical mixing may not be able to explain spectral features arising from photochemical processes.

\begin{figure*}
    \centering
    \includegraphics[angle=0,width=\textwidth]{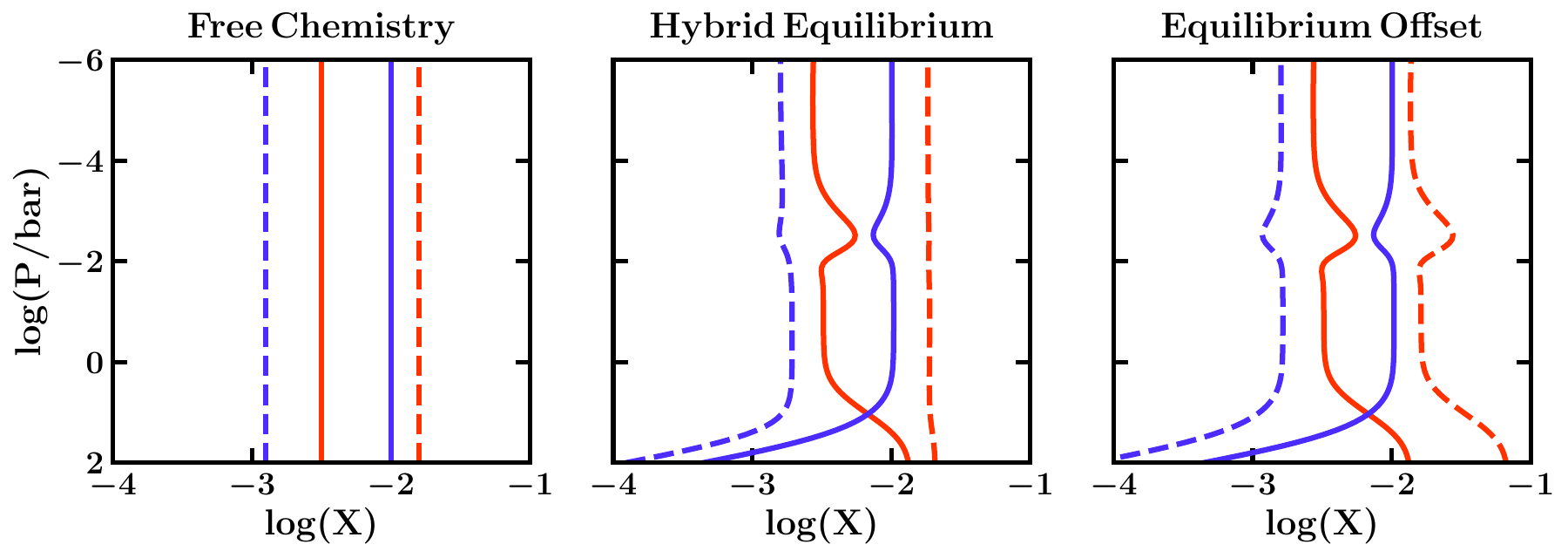}
    \caption[A representation of each of the three approaches to atmospheric chemistry implemented in VIRA.]{A demonstration of each of the three approaches to atmospheric chemistry implemented in VIRA. In all cases, solid lines denote the vertical mixing ratio profiles of H$_2$O and CH$_4$ generated for a particular set of composition parameters. Dashed lines denote the equivalent mixing ratio profiles for a different model, generated for a different set of composition parameters. The comparisons demonstrate how the vertical mixing ratio profiles of the two molecules change as the composition parameters are varied. Changing the mixing ratio free parameters in the free chemistry paradigm results in the constant abundance profiles being shifted. In the hybrid equilibrium case, changing the oxygen and carbon elemental abundances both shifts the mixing ratio profiles and changes their shape. Lastly in the offset equilibrium case, changing the multiplicative offsets for H$_2$O and CH$_4$ results in their non-uniform mixing ratio profiles shifting but maintaining their shapes.}
    \label{vira_fig:composition_approach_comparison}
\end{figure*}

The high precision of JWST observations, which lead to highly precise constraints for atmospheric properties, sets a high bar for the accuracy of atmospheric models used in retrievals. Moreover, the large wavelength coverage achieved by JWST paves the way towards simultaneous detections of numerous chemical species, including hitherto undetected species. Rising to the challenge posed by present and upcoming JWST observations, VIRA implements three complementary approaches to modelling a planet's atmospheric composition, to be used as part of the cascading retrieval architecture. For very high quality observations, the three approaches are intended to be used in sequence over multiple retrievals, with the results obtained at each step being used to inform model choices for subsequent retrievals. Figure \ref{vira_fig:composition_approach_comparison} illustrates how vertical mixing ratios are treated with each of the three approaches.

\subsubsection{Free Chemistry}
\label{vira_subsubsec:free_chemistry}
The first approach to modelling atmospheric composition is a free retrieval, equivalent to the approach of prior AURA-family retrieval frameworks as described in \citet{Pinhas2018}. As mentioned above, this approach treats the mixing ratio of each chemical species as a free parameter, and assumes the same composition at all altitudes.

As each chemical species' mixing ratio is an independent atmospheric parameter, this approach can be readily used to obtain a detection significance for each chemical species. Additionally, thanks to its flexibility, a wide array of chemical species can be readily included and assessed. A decision can then be made whether to include each molecule in subsequent, more advances stages, based on the constraints obtained and their physical plausibility.

In the case of very precise JWST observations, such as for hot giant exoplanets, the assumption of a uniform composition throughout the atmosphere can potentially bias the retrieved abundance constraints. The free chemistry approach is therefore intended to serve as a first assessment, readily determining which species are present and providing indicative abundance estimates. These results inform which species are to be included in the subsequent retrievals employing the two other approaches to atmospheric composition described below. For less precise JWST data, such as those expected to be obtained by observing only a few transits of sub-giant planets, the free chemistry approach is expected to remain a viable standalone analysis pathway.

\subsubsection{Hybrid Equilibrium Chemistry}
\label{vira_subsubsec:hybrid_equilibrium_chemistry}

The second approach to atmospheric composition implemented in VIRA is a hybrid equilibrium chemistry model. In this case, the free parameters presently consist of the ratios of Oxygen, Carbon, Nitrogen and Sulfur to Hydrogen, relative to solar values. For each model evaluation, an equilibrium chemistry calculation is performed to obtain the mixing ratios of a range of molecules, including H$_2$O, CH$_4$, CO, CO$_2$ NH$_3$, HCN, C$_2$H$_2$ and H$_2$S, for every vertical point in the atmosphere. As such, this method considers variations in mixing ratios for different pressures and temperatures throughout the model atmosphere, unlike the free chemistry case. This means that beyond the input elemental abundances, a given forward model's mixing ratio profiles are also influenced by its specific pressure-temperature (P-T) profile.

In order to maintain maximum flexibility, VIRA implements a hybrid approach which allows for additional species to be included without being part of the thermochemical equilibrium calculation. For such species, the retrieval reverts to constant-with-altitude mixing ratios which are individual free parameters in the retrieval. This allows for trace molecules such as SO$_2$ to be included in the model, which are primarily produced by disequilibrium processes such as photochemistry and are not expected to take up a significant amount of the atmospheric elemental budget. Atomic species like Na and K are also included in this manner, as their mixing ratios are expected to be largely constant with altitude.

For maximal robustness, VIRA carries out the equilibrium chemistry calculations using FastChem 2  \citep{Stock2022}. The calculations consider a total of 523 chemical species, of which 495 are molecules and 28 are in elemental form. Any combination of these chemical species can be included in the final model atmosphere, provided their cross-sections are known. FastChem 2 has been validated over the entire pressure range of our model atmosphere and for temperatures between 100-6000~K. We note that while the present implementation uses the elemental ratios of O, C, N and S to H, the framework can be readily expanded in the future to consider any other element as well. As a fallback, VIRA can alternatively use the analytical prescription presented by \citet{Heng2016}, which considers 8 prominent O-, C- and N-bearing species between 500-3000~K.

Through this hybrid equilibrium retrieval approach, we can directly obtain a first estimate for the atmospheric elemental inventory of the observed planet. As such, this method is an expansion on previous work carried out to constrain the elemental inventory of WASP-39~b using a free retrieval approach on early JWST observation \citep{Constantinou2023}.

While this hybrid equlibrium chemistry approach is more flexible than frameworks that only consider the atmospheric metallicity and C/O ratio as free parameters, the requirement for chemical equilibrium can result in a less than ideal fit being imposed. This can occur if a particularly precise constraint is obtained for the abundance of a particular species, e.g. CO$_2$, which in turn necessitates high abundances of a molecule that may be depleted by disequilibrium processes, e.g. CH$_4$. VIRA's third approach detailed below addresses this deficiency.

\subsubsection{Equilibrium Offset Chemistry}
\label{vira_subsubsec:offset_retrievals}

The third approach to atmospheric composition once again considers an equilibrium chemistry calculation. This time, however, elemental abundances remain fixed, while the vertical mixing ratio of each chemical species is enhanced or depleted by a constant multiplicative factor, each of which constitutes a separate free parameter. This approach is therefore a modern equivalent to that followed by \citet{Madhusudhan2009}.

The particular shape of a given species' mixing ratio profile can be affected by the specific choice of elemental abundances. The elemental abundances are therefore set to the median values obtained with the hybrid equilibrium retrieval described above, which are effectively best guesses for the elemental abundances. Additionally, as mixing ratio profiles depend on the specific pressure-temperature structure, the equilibrium calculation is run anew for each forward model. As with the hybrid equilibrium approach, the chemical equilibrium calculation can be carried out using FastChem 2 \citep{Stock2022} or the analytical approach presented by \citet{Heng2016}. Despite the elemental abundances being fixed, the chemical equilibrium calculation is carried out anew for each forward model for its specific P-T profile, as the temperature at each altitude can significantly affect the resulting chemical composition.

With this approach, it is possible to include the effects of disequilibrium chemistry depleting or enhancing specific molecules while at the same time using physically plausible vertical mixing ratio profiles. It effectively treats the equilibrium chemistry vertical mixing ratio profile as a best guess with regard to its shape. The multiplicative offset of each profile effectively models the effects of disequilibrium chemistry, which are non-trivial to compute from first principles, but can result in significant enhancements or depletions of certain species. As noted above, it is possible that disequilibrium effects may bias the prior hybrid equilibrium retrieval towards inaccurate elemental abundances, which are then passed to the equilibrium offset retrievals. In such cases, the individual offsets of each chemical species means this can be compensated for, as demonstrated in Appendix \ref{vira_sec:appendix}. In extreme cases, where all species demonstrate significant offsets indicating highly inaccurate given elemental abundance ratios, the equilibrium offset retrieval can be run again with adjusted elemental ratios.

As with the hybrid equilibrium approach, equilibrium offset retrievals can include additional species as free chemistry parameters. Molecules such as SO$_2$, which are primarily present due disequilibrium processes, are included in this way, since mixing ratio profiles derived from equilibrium chemistry are not indicative of their actual profile shapes.  We also include Na and K as free chemistry parameters, as their vertical mixing ratio profiles are expected to be largely constant with altitude. They are also the only reference for the atmospheric Na and K budget, and will therefore have no deviation from elemental abundances retrieved from hybrid equilibrium retrievals by construction, provided the retrieval does not find a significantly different spectral baseline.

This approach is the most physically accurate of the three methods available to VIRA. It builds on the information obtained from prior retrievals, using the canonical set of gaseous and aerosol species determined from free chemistry retrievals, and the fiducial elemental abundances obtained from hybrid equilibirum retrievals. The equilibrium offset chemistry approach shares the flexibility of the free chemistry approach in being able to characterise atmospheres with significant disequilibrium chemistry effects at play. At the same time, it does not suffer from the potential accuracy problems that may arise when assuming a constant composition with altitude. In the case of highly precise observations, results from all three approaches are assessed to obtain a consistent picture. We note that with this approach, elemental abundances are not conserved when a given molecule's mixing ratio profile is shifted. The elemental abundances of the observed photosphere must therefore be inferred from the constraints in the post-proccessing stage of the retrieval cascade, which is detailed below.

\subsubsection{Higher-Order Atmospheric Constraints and Inferring Disequilibrium}
\label{vira_subsubsec:constraint_postprocessing}

Since atmospheric opacity can vary significantly across the wide wavelength range spanned by JWST observations, transmission spectroscopy may probe various different altitudes of the atmosphere. The altitude that is probed depends on the overall atmospheric opacity at each wavelength, as spectral regions with significant absorption mean that the transition from opaque to transparent, which gives rise to the observed spectrum, happens at higher altitudes, and vice versa. As abundance constraints obtained with JWST can be extremely precise, their interpretation must also consider which region of the atmosphere gives rise to them. VIRA implements additional post-processing functionality to facilitate the interpretation of the retrieved atmospheric compositional constraints.

The first functionality computes the median retrieved mixing ratio profiles for each species present in the model. For retrievals using the hybrid equilibrium or equilibrium offset approach, this is done by considering a large number of samples from the retrieval chain. These samples are used to construct corresponding vertical mixing ratio profiles for the specific abundance and temperature structure parameters. The mixing ratio profiles are subsequently combined to obtain the  16$^{\mathrm{th}}$, 50$^{\mathrm{th}}$ and 84$^{\mathrm{th}}$ percentile contour at each pressure point, which correspond to the lower 1-$\sigma$ contour, median and upper 1-$\sigma$ contour respectively. This approach therefore is equivalent to how median spectral fits and uncertainty contours are generated. Meanwhile, for the free chemistry approach, which assumes a constant composition with altitude, the retrieved mixing ratio profiles can be constructed from the median and 1-$\sigma$ values retrieved for each species' mixing ratio.

The second functionality constructs the median atmospheric pressure probed by the transmission spectrum at each wavelength. This effectively establishes the planetary ``photosphere'' which primarily gives rise to the observed transmission spectrum. This also relies on sampling the retrieval chain, computing a model atmosphere for each sample.The pressure probed by the transmission spectrum is assumed to be where the atmospheric optical depth equals unity. We obtain the median and 1-$\sigma$ contours in a manner similar to the retrieved mixing ratio abundances. Specifically, at each wavelength point, we consider the set of pressure values probed by the computed models, extracting the  16$^{\mathrm{th}}$, 50$^{\mathrm{th}}$ and 84$^{\mathrm{th}}$ percentile values, which we use to construct the median and 1-$\sigma$ contours.

Together, the two functionalities enable a more detailed analysis of the retrieved atmospheric composition.  We note that as our post-processing relies on analysing models that correspond to samples from the retrieval chain, any correlations in the observations (described in section \ref{vira_subsec:niriss_covariance}) are accounted for by default in the computed 1-$\sigma$ contours. For molecules that are expected to be present from equilibrium chemistry such as H$_2$O and CO, these functionalities provide insight into their abundances at specific altitudes. Additionally, for molecules which are the product of disequilibrium processes, e.g. SO$_2$ which is expected to be produced at high altitudes by photochemistry, the $\tau = 1$ pressure near its spectral feature is an estimate of the maximum pressure at which this molecule is probed. By also considering the retrieved P-T profile, it is possible to understand the temperature of the atmospheric region probed by the observations, and the interplay between the temperature and composition profiles. Finally, this analysis enables a rigorous approach to computing the retrieved elemental abundances by considering the molecular abundance profiles solely across the observed photospheric pressures.

\subsection{Atmospheric Aerosol Models}
\label{vira_subsec:nongrey_clouds}

\begin{figure*}
    \centering
    \includegraphics[angle=0,width=\textwidth]{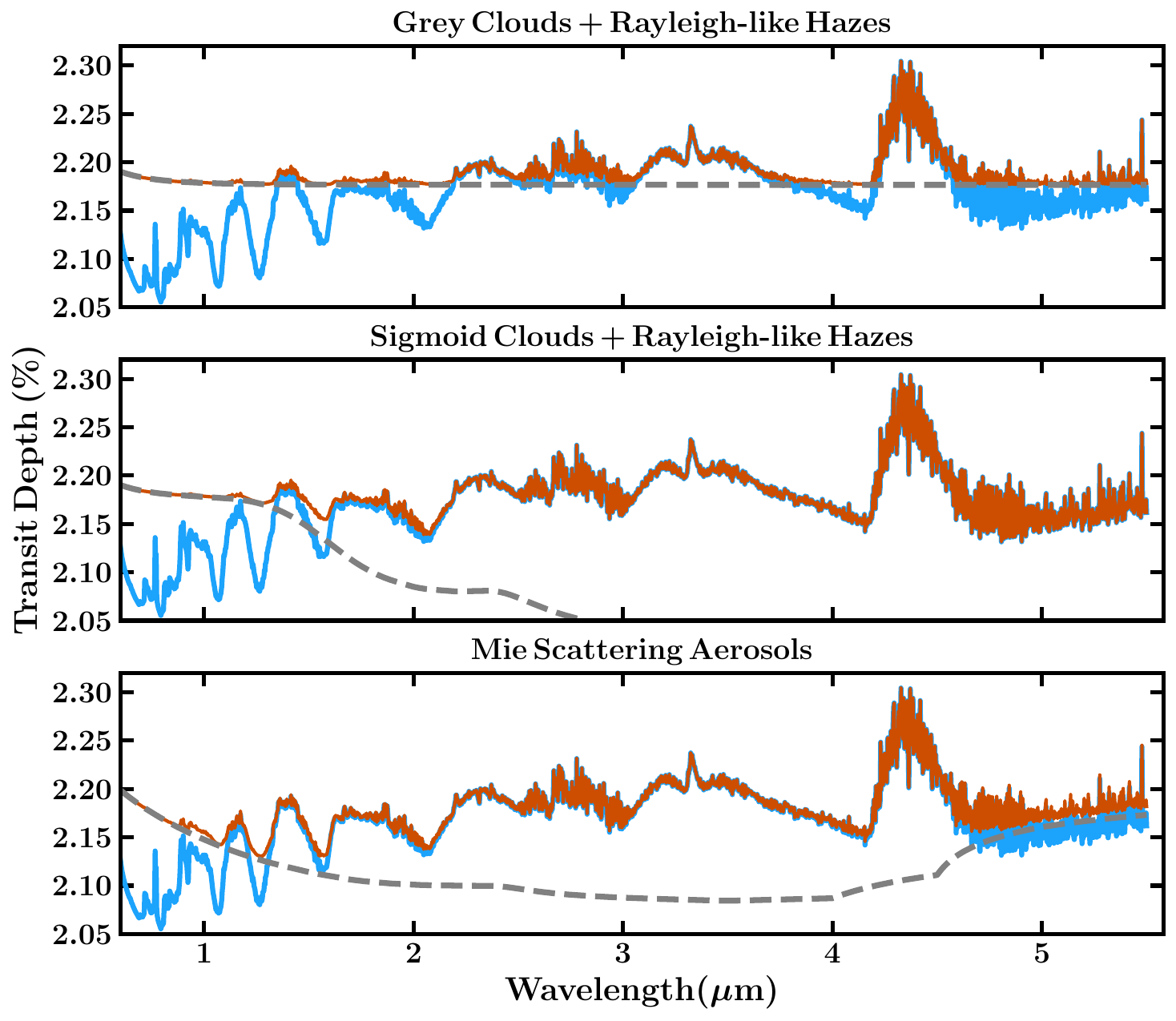}
    \caption[A comparison of the three atmospheric aerosol models available to VIRA, applied to a nominal 10$\times$~solar elemental abundance atmosphere.]{A comparison of the three atmospheric aerosol models available to VIRA, applied to a nominal 10$\times$~solar elemental abundance atmosphere. In all cases, the blue curve corresponds to the spectrum arising from a cloud-free atmosphere and the orange curve to the same terminator atmosphere with 100\% aerosol coverage. The grey dashed line denotes the spectral contributions of atmospheric aerosols. The top panel shows the parametric grey clouds and Rayleigh-like hazes model inherited from prior AURA-family retrievals, with the cloud deck set to 1~mbar and hazes to 100$\times$ enhancement over Rayleigh scattering. The second panel demonstrates general sigmoid cloud approach implemented in VIRA, with $\lambda_\mathrm{Sig} = 1 \mu$m and $w = 4 (\mu \mathrm{m})^{-1}$. Also included are the same 100$\times$ Rayleigh scattering-enhanced hazes as in the grey clouds case. The third panel shows the physical Mie scattering aerosol model, in this case generated for MgSiO$_3$ aerosols with a modal size of 0.01~$\mu$m, mixing ratio of 10$^{-10}$ and full vertical extent.}
    \label{vira_fig:aerosol_comparison}
\end{figure*}

 Accounting for the effects of cloud and haze aerosols is of paramount importance in atmospheric retrieval frameworks, as contributions from clouds and hazes lead to a truncation of absorption features by gaseous species. If such contributions are not modelled appropriately, retrievals may be led to invoke inaccurately low temperatures and/or depleted abundances to explain the smaller than expected features.

While clouds can be approximated as a grey opacity when considering a more limited wavelength range such as achieved by HST, the wavelength range spanned by JWST observations means this assumption may no longer be accurate. Moreover, it has already been shown by prior theoretical and observational works that retrievals on JWST observations can use the variation of an aerosol layer's opacity with wavelength to constrain its properties \citep{Wakeford2015, Pinhas2017, Mai2019, Lacy2020a, Constantinou2023}.

VIRA implements two new approaches to modelling opacity contributions from aerosols and also inherits the parametric clouds/hazes aerosol models used in other AURA-family retrievals \citep{Macdonald2017, Pinhas2018, Welbanks2021, Nixon2022}. The first new approach implemented in VIRA replaces the grey cloud assumption used in the HST era with a parametric opacity model. The second new model is a physical Mie scattering aerosol model, which considers specific aerosol compositions and their physical properties. Specifically, it builds on that used by \citet{Constantinou2023} and includes a larger set of aerosol species, allowing for any number of species to simultaneously be included in the retrieval. The choice of aerosol model will depend on the quality of the available observations, with the first enabling constraints on the specific physical properties of the atmospheric aerosols, and the second allows for maximal generality. A comparison of the extinction contributions from atmospheric aerosols for the three models implemented in VIRA is shown in Figure \ref{vira_fig:aerosol_comparison}.

\subsubsection{Parametric Clouds/Hazes Models}
\label{vira_subsubsec:parametric_cloudsandhazes}
As noted above, VIRA inherits the parametric clouds/hazes models used in other AURA-family retrievals \citep{Macdonald2017, Pinhas2018, Welbanks2021, Nixon2022}. Such models combine spectral contributions from a grey opacity cloud deck and Rayleigh-like scattering hazes above the clouds and are described in detail by \citet{Pinhas2018} and \citet{Welbanks2021}. In summary, the extinction arising from the scattering hazes is described by a modification of H$_2$ Rayleigh scattering, parametrised by $\mathrm{log(}\mathrm{a)}$, the multiplicative enhancement factor, and $\gamma$, the scattering slope. Meanwhile the cloud deck is described by its cloud top pressure $P_\mathrm{c}$. These models allows for only a fraction of the planet's terminator to be covered by clouds and hazes, with AURA \citep{Pinhas2018} relying on a global coverage fraction parameter applying to both clouds and hazes, while Aurora considers three separate coverage fractions for clouds, hazes and the two combined. 

\subsubsection{A General-Purpose Cloud Parametrisation}
\label{vira_subsubsec:sigmoid_clouds}

VIRA implements a new parametric approach to modelling aerosol opacity contributions. This constitutes an evolution of the parametric aerosol models used in prior AURA-family retrieval frameworks described above, which comprise of a patchy grey cloud deck and modified Rayleigh-like scattering above the cloud deck \citep{Macdonald2017, Pinhas2018, Welbanks2021, Nixon2022}.

This evolution is motivated by the effective cross-sections of Mie scattering aerosols below $\sim$10~$\mu$m, as shown by \citet[e.g.][]{Wakeford2015, Pinhas2017}. For such wavelengths, aerosols typically display a near-constant effective cross section at wavelengths around 1~$\mu$m, which decreases at longer wavelengths before reaching a minimum. This decrease varies depending on the specific species and their modal particle size. Following this behaviour, we implement a parametric cloud deck whose contributions to the atmospheric optical depth varies as a sigmoid:

\begin{equation}
\tau_{\mathrm{c}}(\lambda) = \frac{100}{1 + \mathrm{exp}(w(\lambda - \lambda_\mathrm{Sig}))}.
\label{vira_eqn:tau_clouds}
\end{equation}

This sigmoid opacity cloud deck is therefore described by three parameters: $P_\mathrm{c}$, the cloud top pressure, equivalently to prior grey cloud parametrisations; $\lambda_\mathrm{Sig}$, which specifies the centre of the sigmoid and hence the wavelength at which cloud opacity subsides; and $w$, which controls the rate of decrease of the optical depth around $\lambda_\mathrm{Sig}$. The sigmoid cloud model reduces to grey opacity for large $\lambda_\mathrm{Sig}$ and $w$. VIRA can also include modified Rayleigh-like scattering above the sigmoid cloud deck, similarly to other AURA-family frameworks as described above and by \citet{Pinhas2018}. In total, this parametric model consists of 6 free parameters, of which 3, $P_\mathrm{c}$, $\lambda_\mathrm{Sig}$ and $w$, describe the cloud deck, 2, $\mathrm{log}(a)$ and $\gamma$, control the Rayleigh-like hazes and a final parameter, $f_\mathrm{c}$ defines the fraction of the planet's terminator covered by aerosols.

We note that in actuality, the opacity contributions of Mie scattering aerosols may display features in addition to the general decreasing trend.  Depending on their composition, aerosols may display such features at wavelengths beyond $\sim5 \mu$m (e.g. MgSiO$_3$) or at shorter wavelengths (e.g. tholins, H$_2$O). As such, the sigmoid model is appropriate as a diagnostic for the presence non-grey aerosol opacity in a retrieval cascade, justifying the usage of more sophisticated aerosol models in subsequent stages of the cascade. The sigmoid model may also be sufficient on its own in cases where the available observations are not of sufficient quality to resolve more detailed scattering features.

\subsubsection{Inhomogeneous Mie Scattering Aerosols}
\label{vira_subsubsec:mie_aerosols}
In addition to the above parametric approaches, VIRA can include the spectral contributions of atmospheric aerosols using the Mie scattering model described and used by \citet{Constantinou2023}. Here we briefly summarise the model's key aspects.

The Mie scattering aerosol model used in VIRA can include any number of different aerosol species, each parametrised by a separate mixing ratio, defined at the bottom of the model atmosphere. Additionally, the model considers three parameters which universally apply to all aerosols considered. These are the modal particle radius $r_\mathrm{c}$, the terminator coverage fraction $f_\mathrm{c}$ and the aerosol effective scale height, $H_\mathrm{c}$, which is expressed in terms of the atmospheric scale height and controls the rate at which the aerosols' mixing ratios decrease with altitude. 

VIRA relies on the Mie scattering calculation presented by \citet{Pinhas2017}. For each aerosol composition and modal particle size, we use the complex refractive index data compiled by \citet{Wakeford2015} to compute the corresponding effective cross-sections, which considers both the thermalisation of the aerosol particles as well as anisotropic scattering.

As noted above, this approach has already been used by \citet{Constantinou2023} to demonstrate that JWST observation can in principle lead to constraints for aerosol properties, using the first available NIRSpec PRISM observations of WASP-39~b \citep{ERS2023}. For the atmospheric retrievals carried out in Section \ref{vira_sec:WASP-39b_retrievals}, we consider opacity contributions arising from ZnS, MgSiO$_3$ and KCl aerosols, with refractive index data originally obtained from \citet{Querry1987}, \citet{Dorschner1995} and \citet{Palik1998}, respectively.

\subsection{Accounting for Correlated Observations}
\label{vira_subsec:niriss_covariance}
VIRA also implements a more sophisticated treatment of JWST observations themselves when assessing goodness of fit. Specifically, VIRA has the capability of considering the full covariance matrix for a dataset when computing the likelihood of a model. This is particularly important for NIRISS observations, which were recently reported to have significant correlations \citep{Holmberg2023}. As noted by \citet{Holmberg2023}, not accounting for such correlation in NIRISS observations can lead to inaccurate or overly precise estimates for atmospheric parameters.

VIRA includes the covariance matrix of a given dataset by computing a more generalised form of the likelihood function. Specifically, for a set of $N$ transit depth observations $\mathbf{d}$ with a corresponding covariance matric $\mathbf{C}$ and a forward model corresponding to a set of transit depths $\mathbf{m}$ at the same spectral resolution as the observations, which was generated with parameters $\mathbf{\theta}$, the likelihood $\mathcal{L}$ is:

\begin{equation}
\mathcal{L}(\mathbf{d} | \mathbf{\theta}, \mathbf{m}) = \frac{\exp\left( -\frac{1}{2} (\mathbf{d} - \mathbf{m})^\mathrm{T} \mathbf{C}^{-1} (\mathbf{d} - \mathbf{m}) \right)}{\sqrt{ (2 \pi)^N |\mathbf{C}| }},
\end{equation}

where $\mathbf{C}^{-1}$ and $|\mathbf{C}|$ denote the inverse and determinant of the covariance matrix, respectively.

Qualitatively, the covariance matrix captures the extent to which each pair of observations are likely to be affected by the same (if positive) or opposite (if negative) offset due to noise. As a result, the goodness of fit of a given model incurs an additional penalty if it misses the two points in opposite directions, if their corresponding covariance is positive. If their covariance is negative, then the goodness of fit incurs a penalty if the model misses the two points in the same direction, as it is expected that their noise-driven offsets from the true values are in opposite directions. In the case of NIRISS, as shown by \citet{Holmberg2023}, the obtained covariance matrices contains both positive and negative non-diagonal elements, at times with magnitudes comparable to the variances along the diagonal. As such, a na\"ive retrieval that does not account for these correlations can lead to inaccurate estimates for atmospheric parameters and may over-estimate their precisions.

\subsection{Atmospheric Temperature Profile}
\label{vira_subsec:PT_profile}

VIRA is capable of modelling the terminator atmosphere's temperature profile using the 6-parameter framework of \citet{Madhusudhan2009}, as with other AURA-family retrieval frameworks. The parameters consist of $T_\mathrm{0}$, the temperature at the top of the atmosphere, $\alpha_1$ and $\alpha_2$, which control the rate at which the temperature profile varies with temperature, and $P_1$, $P_2$, $P_3$, which set the boundaries between the three atmospheric layers that comprise the model. A detailed description of the temperature profile model is presented by \citet{Pinhas2018} and \citet{Welbanks2021}. VIRA can alternatively treat the temperature profile as an isotherm, parametrised by a single temperature value T$_\mathrm{iso}$.

It has already been demonstrated that transmission spectroscopy observations with JWST can in principle lead retrievals to infer a non-isothermal P-T profile for the terminator \citep{Constantinou2023}. Additionally, the hybrid equilibrium and offset equilibrium chemistry approaches to atmospheric composition detailed in \ref{vira_subsec:cascading_architecture} can result in atmospheric composition being tightly coupled with the temperature profile. This is particularly important for planets with temperatures crossing significant chemical transition points, such as that between CH$_4$ and CO as the primary carbon-carrying molecules in H$_2$-dominated atmospheres. As such, we employ the non-isothermal temperature profile model for all retrievals presented in Section \ref{vira_sec:WASP-39b_retrievals}.

\begin{figure*}
    \centering
    \includegraphics[angle=0,width=\textwidth]{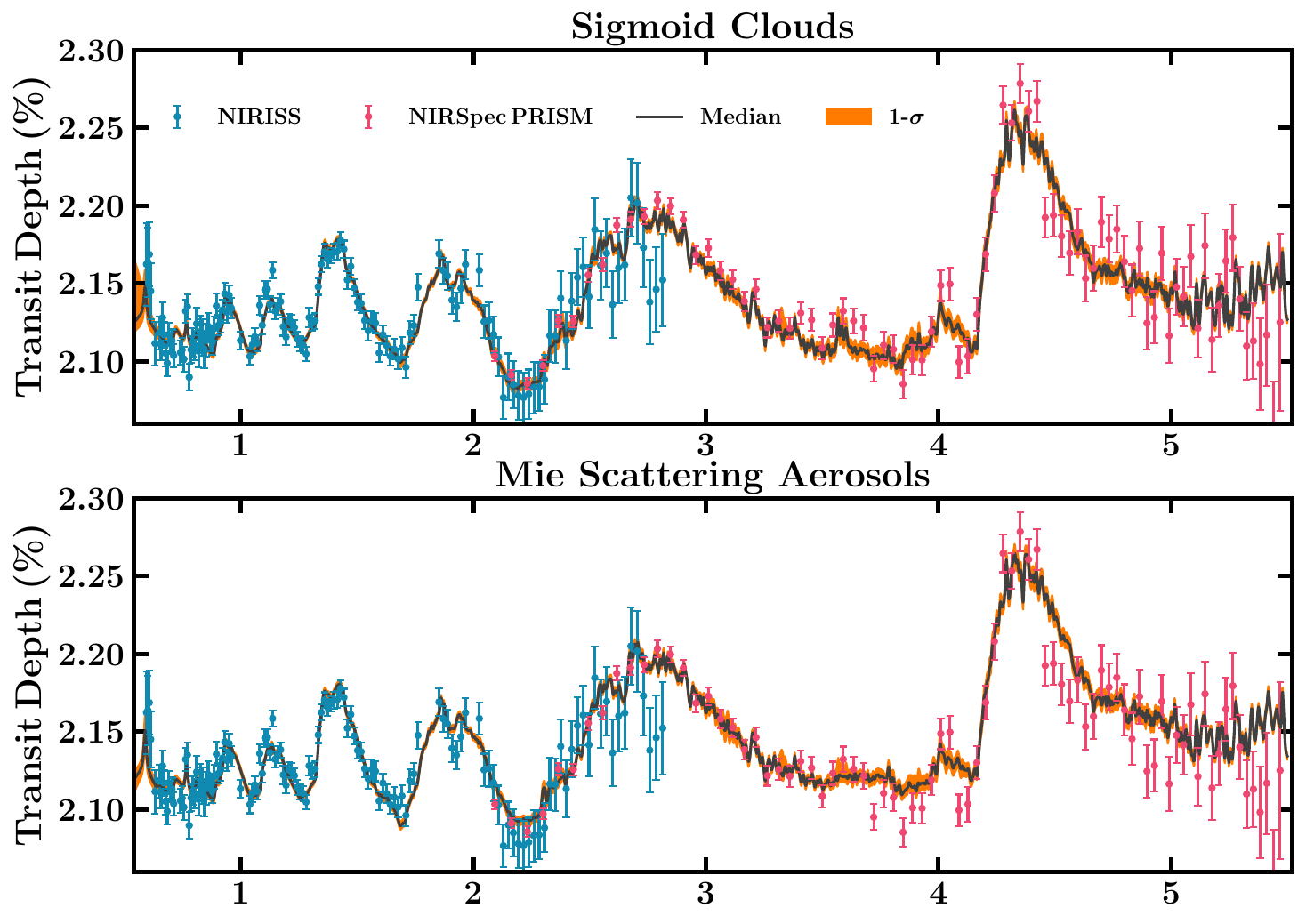}
    \caption{JWST transmission spectroscopy observations of WASP-39~b, obtained with the NIRISS (blue) and NIRSpec PRISM (pink) instruments. Also shown are the median retrieved spectral fit and corresponding 1-$\sigma$ contours obtained with the two free chemistry retrievals presented in section \ref{vira_subsec:free_chemistry_retrievals}. We emphasise that the NIRISS datapoints have significant correlations, which are not conveyed by their representation as independent errorbars.}
    \label{vira_fig:free_spectral_fits}
\end{figure*}

\subsection{Robust Nested Sampling with UltraNest}
\label{vira_subsec:ultranest}

To carry out a robust exploration of the avaiable parameter space, VIRA relies on two implementations of the Nested Sampling algorithm for Bayesian inference \citep{Skilling2004, Skilling2006}. The first is MultiNest \citep{Feroz2009, Feroz2019}, as used in other AURA-family retrieval frameworks. However, Multinest and other similar Nested Sampling implementations may suffer from biases when likelihood contours are non-ellipsoidal \citep{Buchner2016, Nelson2020}.

As noted above, JWST observations enable extremely precise atmospheric constraints for a wide variety of chemical species. At the same time, fully explaining the data requires sophisticated modelling, often with physically motivated parametrisations that can have complex correlations and result in non-ellipsoidal likelihoods. In light of these potential biases, VIRA additionally uses UltraNest \citep{Buchner2021}, which trades computational efficiency for robustness. UltraNest is reserved for the final stages of the retrieval cascade, which rely on the most sophisticated atmospheric models and for very high quality observations, where the the biases from the nested sampling algorithm are potentially important. VIRA maintains the MultiNest implementation for retrievals at earlier stages of the cascade, which are primarily intended to inform the direction of subsequent retrievals. MultiNest also remains the primary workhorse for observations whose quality does not necessitate the full progression down the retrieval cascade.

\begin{figure*}
    \centering
    \includegraphics[angle=0,width=\textwidth]{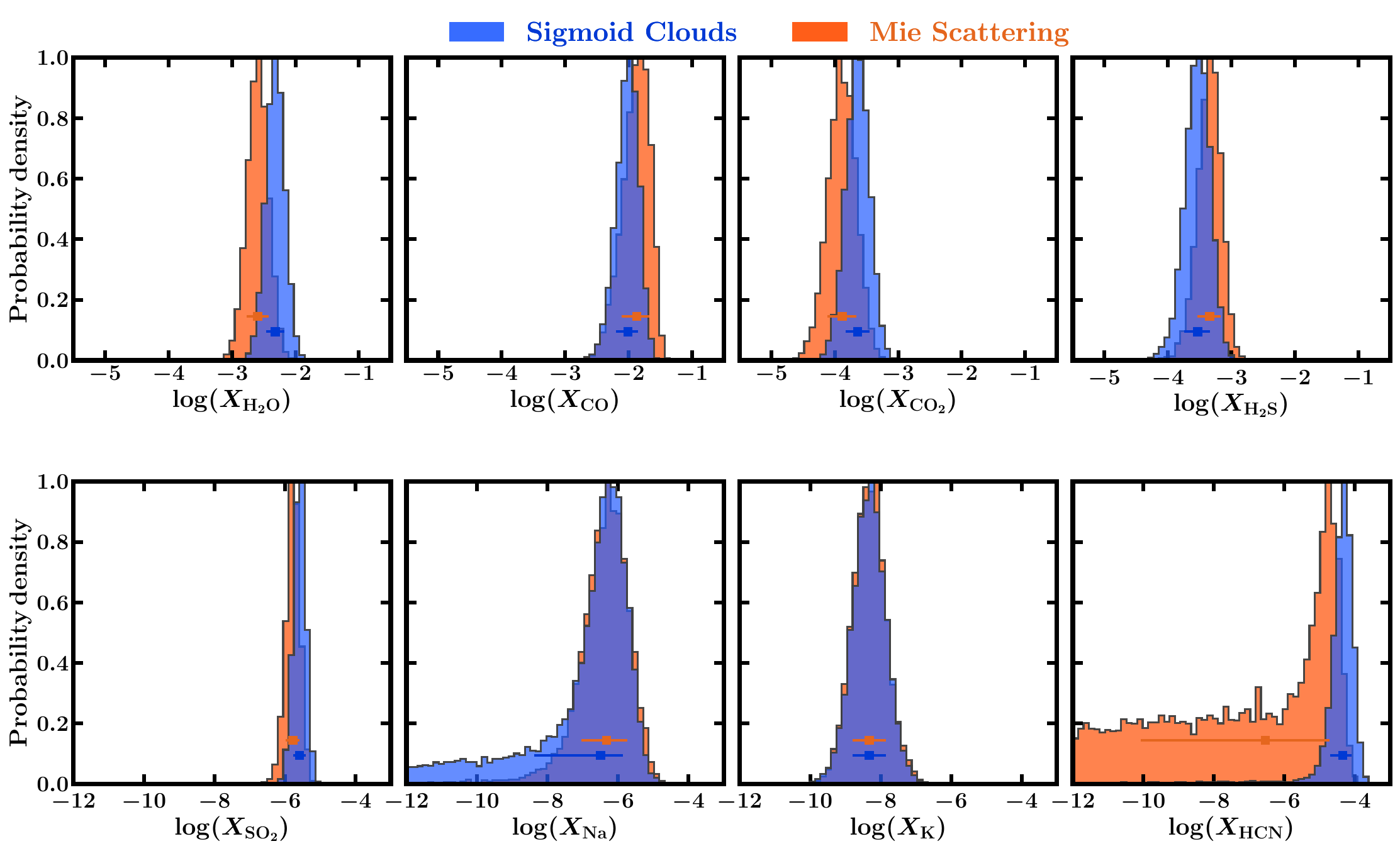}
    \caption{Log-mixing ratio posterior distributions obtained with free chemistry retrievals using the parametric sigmoid cloud model and Mie scattering aerosol model. Horizontal errorbars denote the median retrieved values and corresponding 1-$\sigma$ intervals.}
    \label{vira_fig:free_posteriors}
\end{figure*}

\section{VIRA Retrievals on JWST Observations of WASP-39~b}
\label{vira_sec:WASP-39b_retrievals}

We now demonstrate the capabilities of the VIRA retrieval framework by carrying out a comprehensive analysis of JWST observations of WASP-39~b \citep{Faedi2011}. A Saturn-mass planet (0.3 M$_\mathrm{J}$) with a radius larger than Jupiter's (1.2~R$_\mathrm{J}$), WASP-39~b has an equilibrium temperature of 1170~K, assuming a Bond albedo of zero \citep{Faedi2011, Mancini2018}. It is highly conducive to transmission spectroscopy, thanks to its warm and highly extended atmosphere, as well as the relatively high brightness of its G8 host star (J = 10.8). WASP-39~b has been extensively observed as part of the Transiting Exoplanet Community Early Release Science Program (ID: 1366) using a wide variety of instrument configurations. The extremely high SNR of these observations sets a high bar for model accuracy, making it an ideal case study to demonstrate the capabilities of the VIRA retrieval framework. Appendix \ref{vira_sec:appendix} details an additional retrieval cascade on simulated observations of WASP-39~b, demonstrating the robustness of the approach.

 We specifically consider observations over the 0.6-2.8~$\mu$m range obtained with NIRISS, presented by \citep{Holmberg2023}, and with NIRSpec PRISM, spanning 2-5.5~$\mu$m range, presented by \citet{Rustamkulov2023} and reduced with the Eureka pipeline. We exclude NIRSpec PRISM observations at wavelengths below 2~$\mu$m, in order to avoid observations that may be affected by the detector being saturated or in a non-linear regime. The observations are shown in Figure \ref{vira_fig:free_spectral_fits}. We note that the NIRSpec PRISM dataset used in the present work differs slightly from that considered in \citet{Constantinou2023} which was originally presented by \citep{ERS2023}.

 We note that for all the retrievals presented in this work, we set the planetary radius used in the retrieval based on the white light transit depth from the observations and the measured stellar radius. The retrieval then considers the reference pressure, $P_\mathrm{ref}$, as a free parameter, which is defined as the pressure corresponding to the white-light planetary radius. Since the planetary radius is set to coincide with the averaged transit depth of the transmission spectrum, the retrieved $P_\mathrm{ref}$ value is therefore representative of the averaged effective photosphere pressure.

In the following sections, we present the results obtained for each major stage of the retrieval cascade and how they inform subsequent stages. In all cases, we consider the full covariance matrix of the NIRISS observations, as described in section \ref{vira_subsec:niriss_covariance} and the 6-parameter P-T profile of \citet{Madhusudhan2009} presented in section \ref{vira_subsec:PT_profile}. Additionally, we retrieve for a linear vertical offset between the NIRISS and NIRSpec PRISM data, in order to account for relative differences in transit depth which may arise from different system parameters being assumed in the two reductions. The chemical mixing ratios corresponding to each retrieval's chemical constraints at 10~mbar are summarised in table \ref{tab:retrieved_mixing_ratios}.

\begin{table*}
    
    \centering
    \caption{Retrieved chemical abundances at a pressure of 10~mbar, shown as log-mixing ratios.}
    \begin{tabular}{c|c|c|c|c|c|c|c}
          & H$_2$O & CO & CO$_2$ & H$_2$S & SO$_2$ & Na & K \\
         \hline
         \hline
        \multicolumn{8}{c}{{\it Free Chemistry}} \\
        \hline
        
        Sigmoid Clouds & $-2.32^{+0.15}_{-0.15}$ & $-2.01^{+0.16}_{-0.19}$ & $-3.64^{+0.18}_{-0.19}$ & $-3.53^{+0.19}_{-0.21}$ & $-5.58^{+0.17}_{-0.17}$ & $-6.49^{+0.63}_{-1.87}$ & $-8.32^{+0.47}_{-0.48}$\\[0.2cm]
        Mie Aerosols & $-2.60^{+0.17}_{-0.17}$ & $-1.87^{+0.20}_{-0.24}$  & $-3.89^{+0.22}_{-0.23}$ & $-3.34^{+0.17}_{-0.19}$ & $-5.77^{+0.19}_{-0.21}$ & $-6.32^{+0.58}_{-0.72}$ & $-8.32^{+0.46}_{-0.47}$ \\
        \hline 
        \multicolumn{8}{c}{{\it Hybrid Equilibrium}} \\
        \hline
        Mie Aerosols & $-2.71^{+0.32}_{-0.21}$ & $-2.00^{+0.11}_{-0.11}$ & $-3.76^{+0.44}_{-0.39}$ & $-3.22^{+0.25}_{-0.19}$ & $-5.49^{+0.18}_{-0.16}$ & $-6.40^{+0.93}_{-0.64}$ & $-8.04^{+0.73}_{-0.43}$ \\
        \hline
        \multicolumn{8}{c}{{\it Equilibrium Offset}} \\
        \hline
        MultiNest & $-2.60^{+0.12}_{-0.12}$ & $-2.07^{+0.17}_{-0.19}$  & $-3.69^{+0.19}_{-0.19}$ & $-3.24^{+0.17}_{-0.19}$ & $-5.69^{+0.16}_{-0.18}$ & $-5.66^{+0.51}_{-0.64}$ & $-7.68^{+0.41}_{-0.47}$ \\[0.2cm] 
        UltraNest & $-2.57^{+0.13}_{-0.13}$ & $-2.00^{+0.16}_{-0.20}$ & $-3.69^{+0.22}_{-0.21}$ & $-3.15^{+0.13}_{-0.13}$ & $-5.71^{+0.19}_{-0.23}$ & $-5.37^{+0.31}_{-0.37}$ & $-7.51^{+0.32}_{-0.37}$ \\ 
        \hline
        
    \end{tabular}
    
    \label{tab:retrieved_mixing_ratios}
\end{table*}

\subsection{Prior Inferences with JWST Observations}
\label{vira_subsection:prior_w39b_inference}
The JWST observations of WASP-39~b have already been analysed with various degrees of sophistication and led to several important atmospheric detections and inferences. The first observations presented for WASP-39~b, obtained with NIRSpec PRISM and spanning the 3-5~$\mu$m range, revealed a highly prominent absorption feature attributed to CO$_2$ \citep{ERS2023}, the first such detection made for an exoplanet atmosphere. These observations additionally displays an additional absorption features blueward of the CO$_2$ feature, at $\sim$4~$\mu$m, which could not be readily explained, highlighting the immense discovery space enabled by JWST. The data was also indicative of a 10$\times$~solar atmospheric metallicity, deduced via a forward model comparison.

Subsequent observations obtained with NIRISS \citep{Doyon2012, Feinstein2023}, NIRCam F322W2 \citep{Rieke2005, Ahrer2023} and NIRSpec \citep{Birkmann2014} G395H \citep{Alderson2023}, as well the complete NIRSpec PRISM dataset extending down to $\sim$1~$\mu$m \citep{Rustamkulov2023}, have confirmed the somewhat super-solar metallicity of the atmosphere, again deduced via comparison to forward atmospheric models. Moreover, \citet{Alderson2023} and \citet{Rustamkulov2023} found that the previously unexplained feature at $\sim$4~$\mu$m is due to SO$_2$, which is a product of photochemical processes \citep{Tsai2023}. NIRISS observations meanwhile confirmed HST-era detections of H$_2$O, as well as the presence of clouds.

The first retrieval analysis was carried out by \citet{Constantinou2023}, who considered the 3-5~$\mu$m NIRSpec PRISM observations presented by \citet{ERS2023}, combined with prior HST WFC3 observations presented by \citet{Wakeford2018}. The presence of H$_2$O, CO$_2$, CO and SO$_2$ was confirmed via robust Bayesian model comparison, with additional tentative indications of H$_2$S. Using the retrieved abundance constraints, it was deduced that the atmosphere of WASP-39~b has elemental O, C and S enrichments relative to stellar values that are consistent with the C enrichment of Saturn relative to the sun. Moreover, in certain cases it was found that the observations were best explained by models that include significant opacity from Mie scattering aerosols.

Lastly, \citet{Niraula2023} carried out another retrieval study, using the error-weighted average of all reductions of the NIRSpec G395H observations presented by \citet{Alderson2023}. Across multiple retrievals using different molecular linelists and approaches to the temperature structure, H$_2$O was constrained to mixing ratios of $\sim$5\%, while H$_2$S was constrained to $\sim$1-2\% and SO$_2$ to $\sim$20-30~ppm. These constraints indicate a significantly higher atmospheric metallicity that those obtained by prior studies.

\subsection{Free Chemistry Retrievals}
\label{vira_subsec:free_chemistry_retrievals}

We begin by considering free chemistry retrievals. The first objective of this stage is to assess which atomic and molecular features are present in the transmission spectrum, and should therefore be considered in subsequent retrievals. The observations, shown in Figure \ref{vira_fig:free_spectral_fits}, display numerous absorption features across the wavelength range. The most prominent of those features have already been attributed to H$_2$O, CO$_2$ and SO$_2$ by prior works, with secondary contributions from CO and potentially H$_2$S. We consider these species, as well as several other in the first step of our retrieval cascade, seeking to establish a canonical set of chemical species to use in subsequent retrievals.

The second objective of this stage of the cascade is to establish the presence and nature of atmospheric opacity contributions from aerosols. The observations we consider span a wavelength range between $\sim$0.6-5.5~$\mu$m. Such a wavelength range is large enough to capture significant variation in the opacity contributions from aerosols due to Mie scattering. Additionally, prior works have found that JWST observations of WASP-39~b are indicative of non-grey cloud opacity \citep{Feinstein2023, Constantinou2023}. As such, we do not use the parametric clouds/hazes model VIRA inherits from prior AURA-family frameworks discussed in section \ref{vira_subsubsec:parametric_cloudsandhazes}, and begin our retrieval cascade using the sigmoid clouds model described in section \ref{vira_subsec:nongrey_clouds}. As discussed above, the sigmoid clouds model can readily indicate the presence of non-grey clouds and can be reduced to the parametric grey clouds and Rayleigh-like hazes model for large $\lambda_\mathrm{Sig}$ and $w$. Informed by our constraints, we subsequently consider a retrieval which replaces the sigmoid clouds with more physically accurate Mie scattering aerosols. As with gaseous species, we consider a wide range of possible aerosol species before arriving at a canonical set.

\begin{figure*}
    \centering
    \includegraphics[angle=0,width=\textwidth]{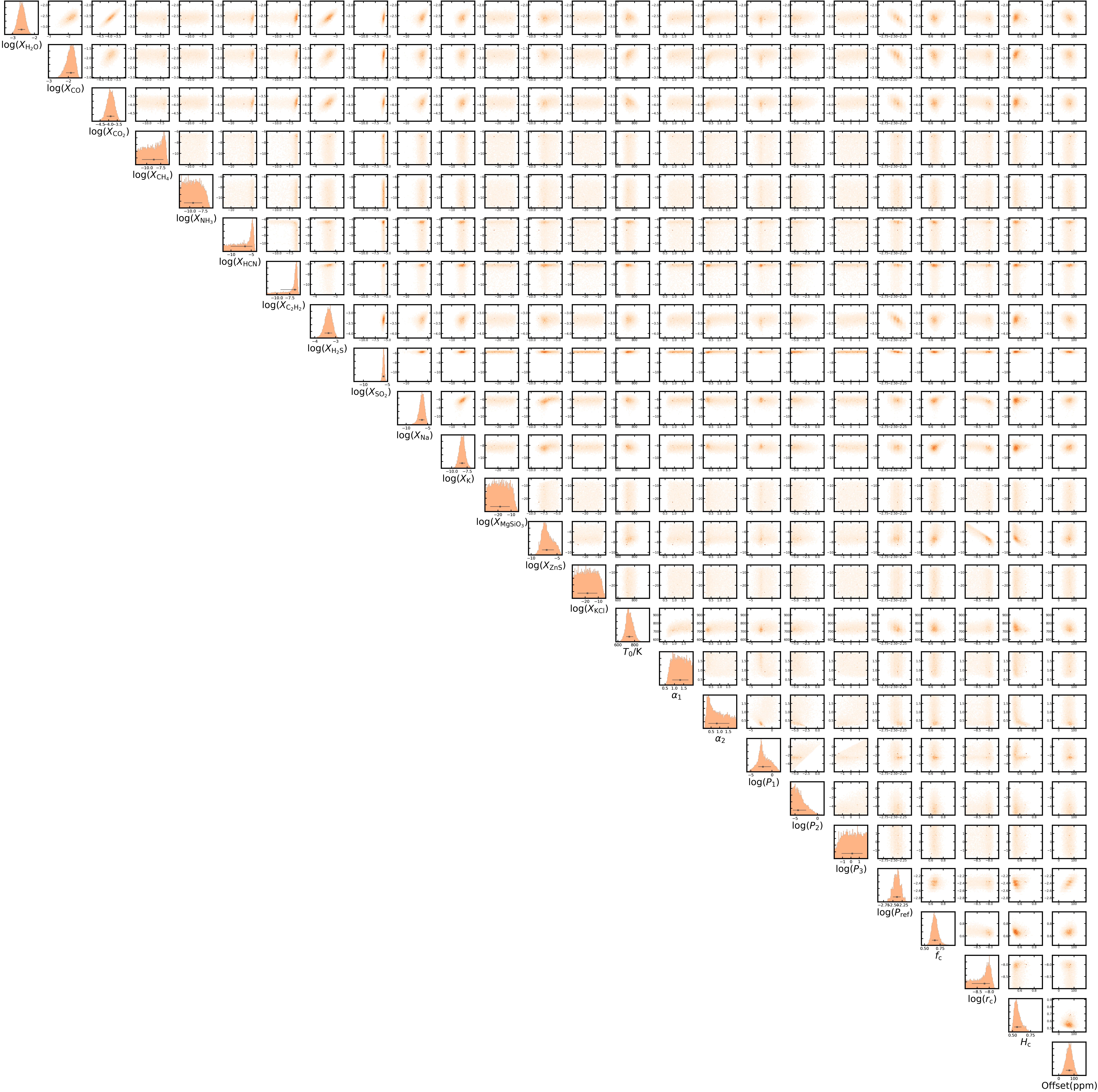}
    \caption{The posterior probability distribution obtained by the free chemistry retrieval using Mie scattering aerosols presented in section 
    \ref{vira_subsubsec:free_miescattering_ret}.}
    \label{vira_fig:free_mie_corner}
\end{figure*}

\subsubsection{Sigmoid Clouds and Choosing a Canonical Set of Chemical Species}
\label{vira_subsubsec:free_sigmoid_ret}
We begin by considering a large number of possible species that may be present in the atmosphere of WASP-39~b, based on the prior detections discussed in section \ref{vira_subsection:prior_w39b_inference} and expectations from thermochemical equilibrium calculations \citep{Burrows&Sharp1999, Lodders&Fegley2002, Zahnle2009, Madhusudhan2012, Moses2013b}. Based on these results, we arrive at a canonical set of chemical species for which we calculate the absorption cross-sections from their line lists: H$_2$O \citep{Barber2006, Rothman2010}, CO \citep{Li2015}, CH$_4$ \citep{Yurchenko2014}, NH$_3$ \citep{Yurchenko2011}, CO$_2$ \citep{Tashkun2015}, HCN \citep{Barber2014}, C$_2$H$_2$ \citep{Chubb2020}, H$_2$S \citep{Azzam2016, Chubb2018}, SO$_2$ \citep{Underwood2016}, Na \citep{Allard2019} and K \citep{Allard2016}. We note that all VIRA retrievals include opacity contributions from H$_2$-H$_2$ and H$_2$-He collision induced absorption \citep{Borysow1988, Orton2007, Abel2011, Richard2012}.

We include H$_2$O and CO as both have already been detected in the atmosphere of WASP-39~b and our present retrieval also constrains their mixing ratios. They are generally expected to be the primary oxygen-carrying molecules in hot, H$_2$-dominated atmospheres, while H$_2$O alone is the dominant oxygen bearer at low temperatures. Additionally, CO along with CH$_4$ are the primary carbon carriers at higher and lower temperatures, respectively. Similarly, CO$_2$ is included as it has been detected at extremely high confidence and our retrieval is able to constrain its mixing ratio. It is also an important marker molecule, with its mixing ratio being extremely sensitive to metallicity \citep{Lodders&Fegley2002, Madhusudhan2011a}. We include NH$_3$ as it is expected to be the primary nitrogen carrier at lower temperatures. At higher temperatures, nitrogen goes to N$_2$, which does not have a significant cross-section in the infrared and is therefore not included in the model. HCN is included as our free chemistry retrieval produces constraints for its mixing ratio, and is an important indicator for high C/O ratios \citep{Madhusudhan2012}. We also include C$_2$H$_2$, as it is another marker molecule for high C/O ratios. H$_2$S is expected to be the primary sulfur-carrying molecule and our retrievals are able to constrain its mixing ratio. We include SO$_2$, an expected product of photochemistry \citep{Zahnle2009, Wang2017, Hobbs2021, Polman2022}, again obtaining constraints for its mixing ratio. Lastly, Na and K are included as they were previously inferred with pre-JWST observations \citep[e.g.][]{Fischer2016, Wakeford2018, Kirk2019, Welbanks2019b} and our retrievals are able to constrain their mixing ratios.

The retrieval achieves a good fit to the observations, as shown in Figure \ref{vira_fig:free_spectral_fits}, and constrains the posterior distributions for a number of chemical species, as shown in Figure \ref{vira_fig:free_posteriors}. We obtain a log-mixing ratio estimate for H$_2$O of $-2.32^{+0.15}_{-0.15}$, driven to a great extent by the several H$_2$O absorption features within the NIRISS wavelength range. Our retrieval also estimates the log-mixing ratio of CO to be $-2.01^{+0.16}_{-0.19}$, which along with H$_2$O provides the atmospheric opacity to explain the redmost data. CO$_2$ is constrained to a log-mixing ratio of $-3.64^{+0.18}_{-0.19}$, primarily driven by the highly prominent absorption feature at $\sim$4.5~$\mu$m, while SO$_2$ is constrained to $-5.58^{+0.17}_{-0.17}$ as a result of its absorption feature at $\sim$4~$\mu$m. Na and K are also constrained, to $-6.49^{+0.63}_{-1.87}$ and $-8.32^{+0.47}_{-0.48}$, respectively, thanks to NIRISS observations encompassing the entire K absorption peak, and the pressure-broadened wing of the Na feature. Lastly, our retrieval obtains constraints for H$_2$S and HCN through their minor contributions to the spectrum at several points. H$_2$S is found to have a log-mixing ratio of $-3.53^{+0.19}_{-0.21}$, while HCN is constrained to $-4.34^{+0.25}_{-0.26}$.

The retrieval constrains the sigmoid cloud parameters to values indicating a significant deviation from grey opacity. Specifically, the mid-point of the sigmoid, $\lambda_\mathrm{Sig}$ is constrained to $1.39^{+0.07}_{-0.11} \mu\mathrm{m}$, while the decay width parameter, $w$ is constrained to $7.15^{+1.74}_{-1.76} (\mu\mathrm{m})^{-1}$. As a result, the opacity contributions from the sigmoid clouds are significant and grey-like between $\sim$0.6-1.5~$\mu$m, start to significantly decrease near the spectral trough at $\sim$1.7$\mu$m and are minimal from the trough at $\sim$2.2~$\mu$m and beyond.

Our model also includes Rayleigh-like haze scattering. While our retrievals do not produce any meaningful constraints for their parameters, their effect can be seen in the 1- and  2-$\sigma$ contours near the Na feature at $\sim$0.6~$\mu$m in Figure \ref{vira_fig:free_spectral_fits}. They effectively permit a near-equivalent alternative explanation of the upward direction of the bluemost NIRISS datapoints, whereby they are the result of significantly enhanced haze scattering rather than the pressure-broadened wings of the Na doublet. Due to this partial degeneracy, the retrieval obtains a lower 1-$\sigma$ interval for the Na log-mixing ratio that is almost 2~dex large.

Lastly, our retrievals obtain a constraint for the log-mixing ratio of HCN. Under chemical equilibrium, such a mixing ratio would correspond to a significantly super-solar C/O ratio, which is not borne out by our constraints for H$_2$O and CO, which indicate C/O ratio consistent with solar. As this is an early stage in our retrieval cascade, the HCN constraint is not taken as definitive.

\begin{figure*}
    \centering
    \includegraphics[angle=0,width=\textwidth]{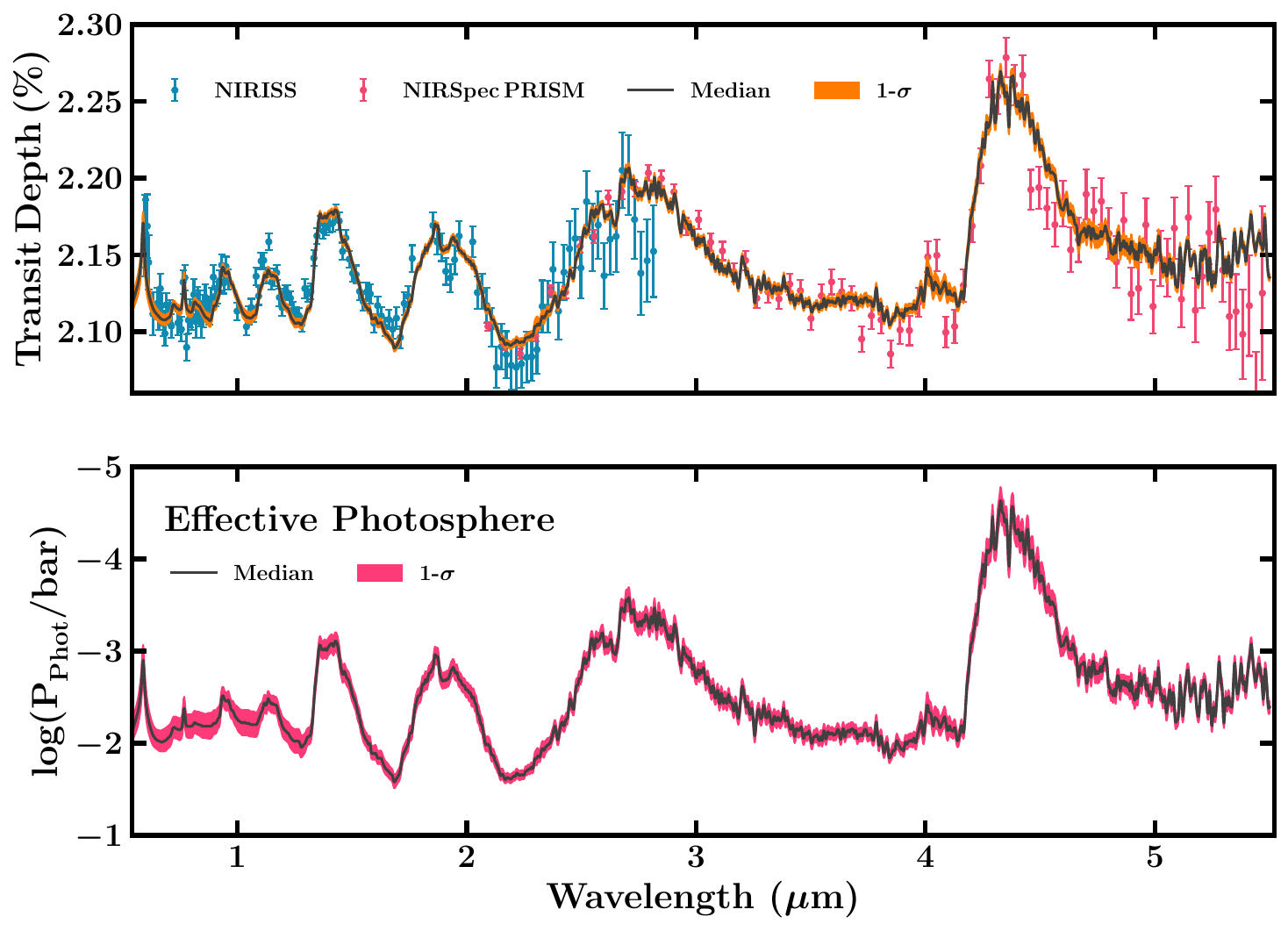}
    \caption{{\bf Top:} The retrieved spectral fit obtained with the equilibrium offset retrieval presented in section \ref{vira_subsec:equilibrium_offset_rets}. The black line denotes the median retrieved fit, while the orange contour denotes the corresponding 1-$\sigma$ interval. Blue and red errorbars denote the NIRISS and NIRSpec PRISM observations. {\bf Bottom:} The median retrieved photosphere pressure (black line) and corresponding 1-$\sigma$ interval (red shaded region) for the same retrieval. The photosphere pressure is defined as the pressure at which a given atmospheric model's optical depth is equal to 1. Specifically, the effective photosphere shown corresponds to the cloudy fraction of the terminator atmosphere, chosen as a conservative estimate. }
    \label{vira_fig:eq_offset_spectrum_photosphere}
\end{figure*}

\subsubsection{Mie Scattering Aerosols}
\label{vira_subsubsec:free_miescattering_ret}

Having established our canonical set of chemical species and that non-grey clouds play a significant role in the interpretation of the data, we now consider free chemistry retrievals using our physical Mie scattering aerosol model. As noted in section \ref{vira_subsec:nongrey_clouds}, our model can consider a wide array of aerosol compositions. We therefore carry out a similar exercise to that done above for gaseous species, assessing a range of potential aerosol compositions to arrive at a canonical set motivated by our retrieved constraints and physical plausibility \citep[e.g.][]{Morley2013}.

Our final Mie scattering model consists of aerosols of MgSiO$_3$, KCl and ZnS. The condensation curves of KCl and ZnS \citep{Morley2013} intersect with the retrieved P-T profile of our prior sigmoid cloud retrievals. Our exploratory retrievals obtain constraints for the mixing ratio of ZnS aerosols, motivating their inclusion. We additionally include MgSiO$_3$ in our canonical model, as prior retrievals on a subset of the present observations found that under certain conditions, there are nominal indications of MgSiO$_3$ aersols being present \citep{Constantinou2023}. While MgSiO$_3$ aerosols condense at higher temperatures than those retrieved for the observable atmosphere above, they may still be present by condensing at lower altitudes and being transported by advection. 

As seen in Figure \ref{vira_fig:free_spectral_fits}, this retrieval obtains a good fit to the observations, which visually does not differ significantly from that obtained by the sigmoid clouds retrieval. The complete posterior distribution obtained for all model parameters is shown in Figure \ref{vira_fig:free_mie_corner}. We retrieve constraints for the abundance of ZnS aerosols, while those of MgSiO$_3$ and KCl remain largely unconstrained, indicating that ZnS aerosols are specifically preferred to explain the data. We obtain a log-mixing ratio of $-6.97^{+1.49}_{-0.87}$, with a corresponding modal particle size $\mathrm{log}(r_\mathrm{c} / \mu \mathrm{m}) = -2.21^{+0.25}_{-0.53}$, effective scale factor $H_\mathrm{c} = 0.55^{+0.04}_{-0.03}$ and terminator coverage fraction $f_\mathrm{c} = 0.64^{+0.06}_{-0.05}$.

Our retrieval obtains mixing ratio constraints for the same gaseous species as with the sigmoid cloud model, as shown in Figure \ref{vira_fig:free_posteriors}. There are however some differences, which are all within 1-$\sigma$ of the constraints obtained above. We obtain a slightly lower H$_2$O log-mixing ratio constraint of $-2.60^{+0.17}_{-0.17}$, which is within 1-$\sigma$ of the equivalent constraint obtained with sigmoid clouds. We also obtain a slightly higher CO log-mixing ratio estimate at $-1.87^{+0.20}_{-0.24}$. Additionally, the log-mixing ratio of CO$_2$ is constrained to $-3.89^{+0.22}_{-0.23}$, and that of SO$_2$ to $-5.77^{+0.19}_{-0.21}$. Additionally, H$_2$S is found at a log-mixing ratio of $-3.34^{+0.17}_{-0.19}$, while Na and K are constrained to $-6.32^{+0.58}_{-0.72}$ and $-8.32^{+0.46}_{-0.47}$.

Our retrieval also constrains the log-mixing ratio of HCN to $-4.86^{+0.39}_{-2.94}$. Notably, the corresponding posterior distribution has a significant ``tail'' towards low abundances, as can be seen in Figure \ref{vira_fig:free_mie_corner}, indicating that models without significant opacity contributions from HCN are also permissible. This is in contrast to our findings with sigmoid clouds, where no such tail was obtained. A similar posterior distribution is obtained for C$_2$H$_2$, displaying a prominent peak and significant spread to low mixing ratios. The retrieved Na log-mixing ratio posterior meanwhile does not have any spread towards low abundances, unlike that obtained with sigmoid clouds. This is due to the more restrictive nature of our physical Mie scattering aerosol model, as the ZnS aerosols favoured by the rest of the data do not have a strong enough scattering slope at short wavelengths to be a competing explanation for the Na doublet wing feature.

The retrieved P-T profile and the corresponding photospheric pressures are shown in Figure \ref{vira_fig:retrieved_Xprofiles_PT}. We find that the retrieved P-T profile is marginally consistent with an isotherm to within 1-$\sigma$ in the observed photospheric pressure range. Specifically, at the highest (upper 2-$\sigma$) altitude of the photosphere, the retrieved temperature is between 710-800~K at the 1-$\sigma$ level. Meanwhile at the lowest altitude of the photosphere, the retrieved temperature ranges between 800-960~K.

Our findings from our free chemistry retrievals show the diversity of atmospheric opacity contributions both from gaseous species as well as aerosols. They also highlight how minor differences between models, such as the atmospheric aerosol model used here, can lead to notably different inferences, particularly for chemical species that are at the limit of observability. Moreover, in the era of highly precise JWST observations, such differences do not manifest in marked differences in the retrieved spectral fits. We therefore emphasise that the mixing ratio constraints obtained here are not definitive. As the retrieved atmospheric constraints are extremely precise, often reaching mixing ratio precisions of $\sim$0.2~dex, it is necessary to continue through our retrieval cascade towards more refined atmospheric models to ensure the retrieved estimates are accurate. Moreover, such precisions are expected to enable more nuanced constraints for aspects such as the vertical mixing ratio profile of each chemical species.

\subsection{Hybrid Equilibrium Retrievals}
\label{vira_subsec:hybrid_equilibrium}

The next step in our cascade towards more physically accurate atmospheric models is to consider atmospheric compositions that are not uniform in altitude. As a first step, we carry out retrievals using VIRA's hybrid equilibrium chemistry approach, as described in section \ref{vira_subsec:cascading_architecture}. We proceed with the canonical sets of gaseous and aerosol species determined in \ref{vira_subsec:free_chemistry_retrievals}. Of those gaseous species, we include a subset in the equilibrium chemistry calculation: H$_2$O, CO, CH$_4$, NH$_3$, CO$_2$, HCN, C$_2$H$_2$ and H$_2$S. The abundances of the remaining species in our canonical gaseous species set, i.e., SO$_2$, Na and K are treated as free log-mixing ratio parameters with constant vertical mixing ratios. The Mie scattering aerosol model remains the same as before, consisting of ZnS, KCl and MgSiO$_3$ aerosols.

The purpose of this stage of the cascade is to obtain a first look at the elemental inventories in the atmosphere, which will be used as fiducial values for our subsequent equilibrium offset stage. Additionally, these retrievals will also give insight to the potential coupling between the temperature profile and the atmospheric composition. This is especially significant as the equilibrium temperature of WASP-39~b indicates that it is at a transitional point between temperate atmospheres, where CH$_4$ dominates that atmospheric carbon budget, and hot atmospheres, with CO dominating the carbon budget and much of the oxygen budget as well.

\begin{figure*}
    \centering
    
    \includegraphics[angle=0,width=0.95\textwidth]{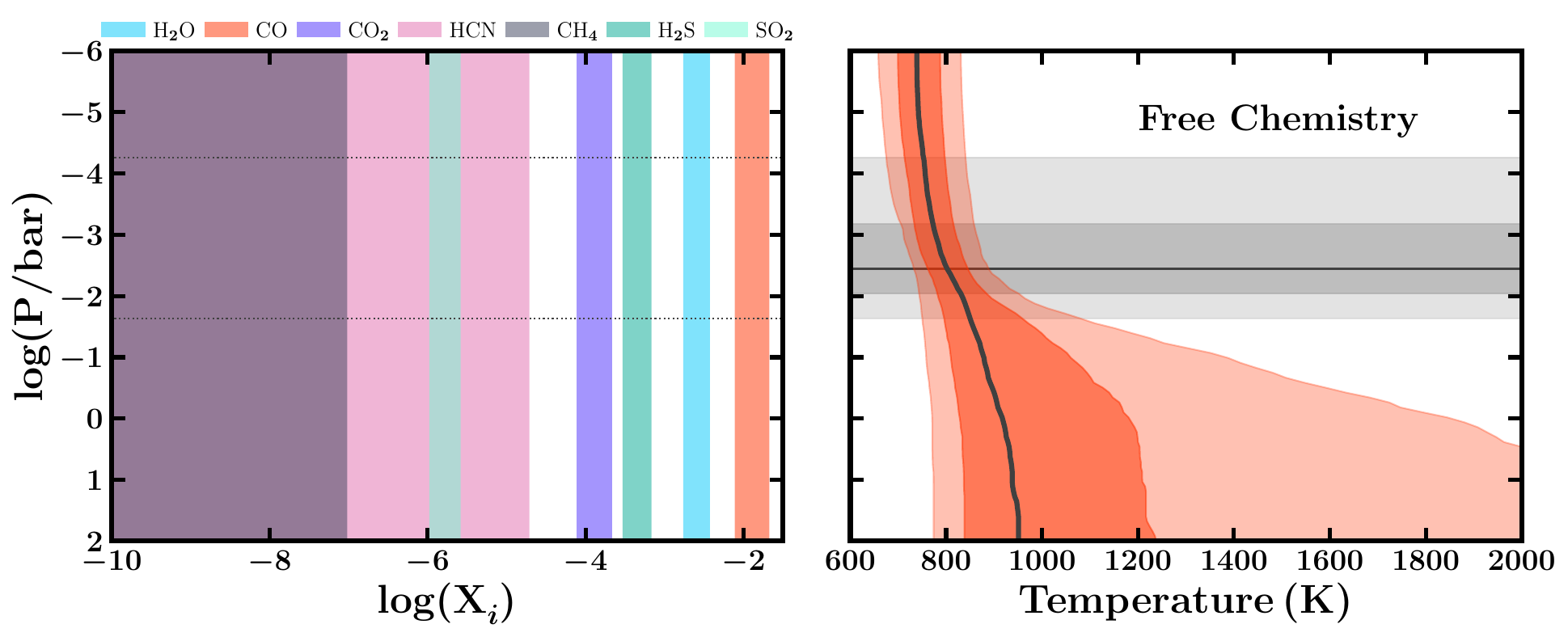}
    \includegraphics[angle=0,width=0.95\textwidth]{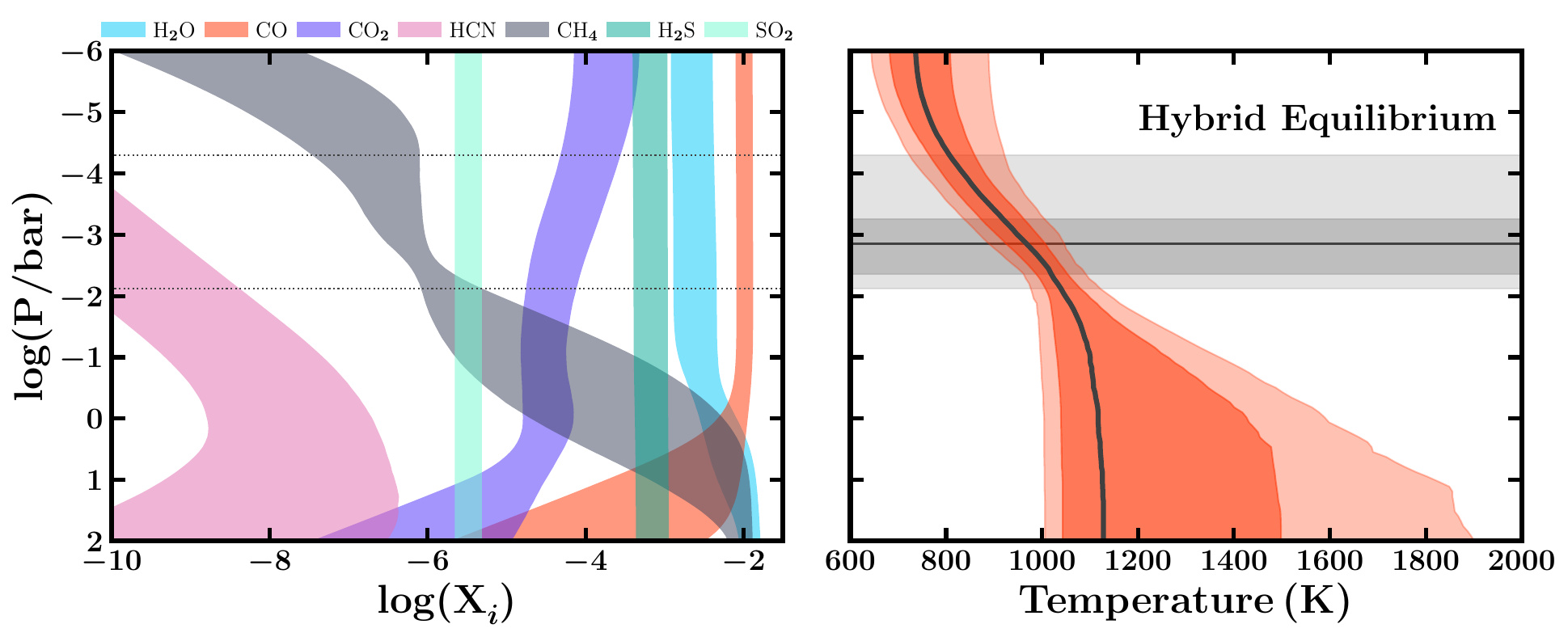}
    \includegraphics[angle=0,width=0.95\textwidth]{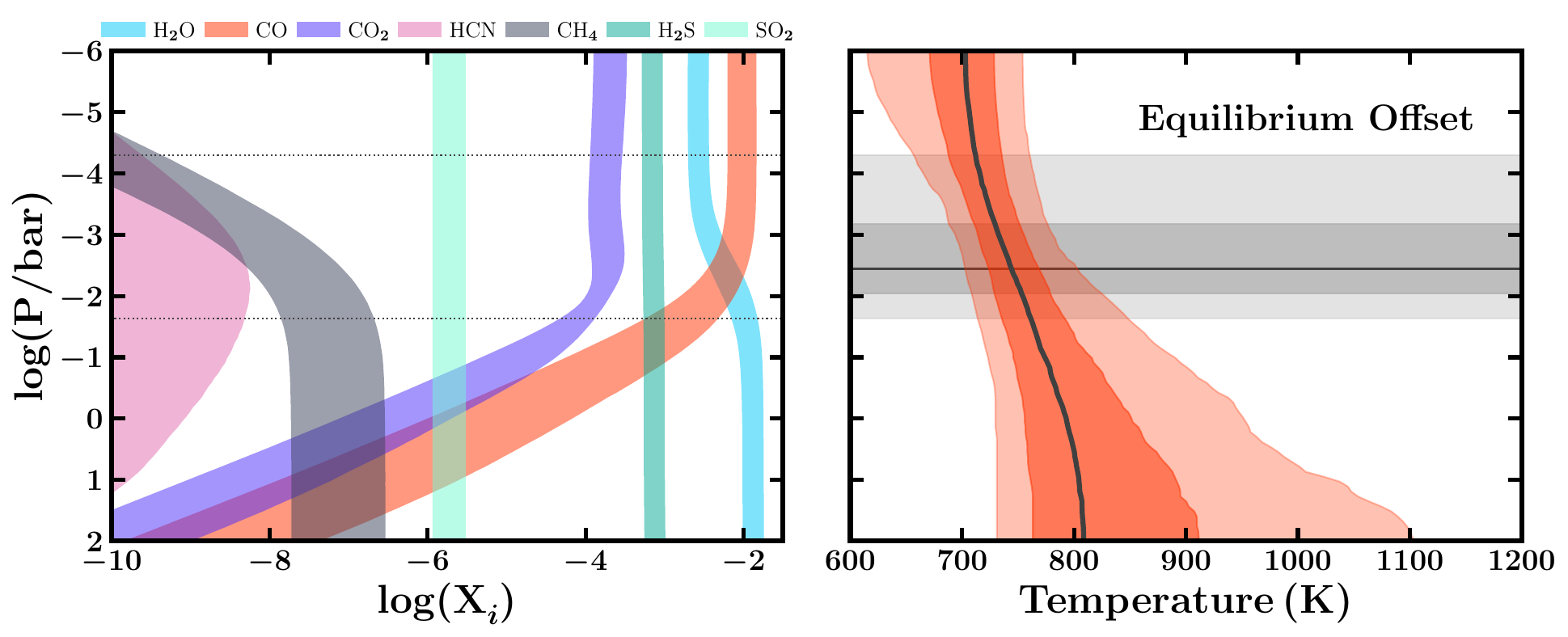}
    \caption{{\bf Left:} Vertical mixing ratio profiles retrieved with the free chemistry, hybrid equilibrium and equilibrium offset retrievals presented in sections \ref{vira_subsec:free_chemistry_retrievals}, \ref{vira_subsec:hybrid_equilibrium} and \ref{vira_subsec:equilibrium_offset_rets}, respectively. The shaded regions denote the 1-$\sigma$ interval for each chemical species' volume mixing ratio. Horizontal dotted lines denote the 2-$\sigma$ contour of the retrieved photospheric pressure across the wavelength range for each retrieval.  {\bf Right:} The retrieved P-T profile for the same retrievals. The black line denotes the median value, while darker and lighter shaded regions denote the 1- and 2-$\sigma$ regions. The horizontal black line denotes the median retrieved photospheric pressure, while darker and lighter grey regions show the corresponding 1- and 2-$\sigma$ contours.}

    \label{vira_fig:retrieved_Xprofiles_PT}
\end{figure*}

\begin{figure*}
    \centering
    \includegraphics[angle=0,width=\textwidth]{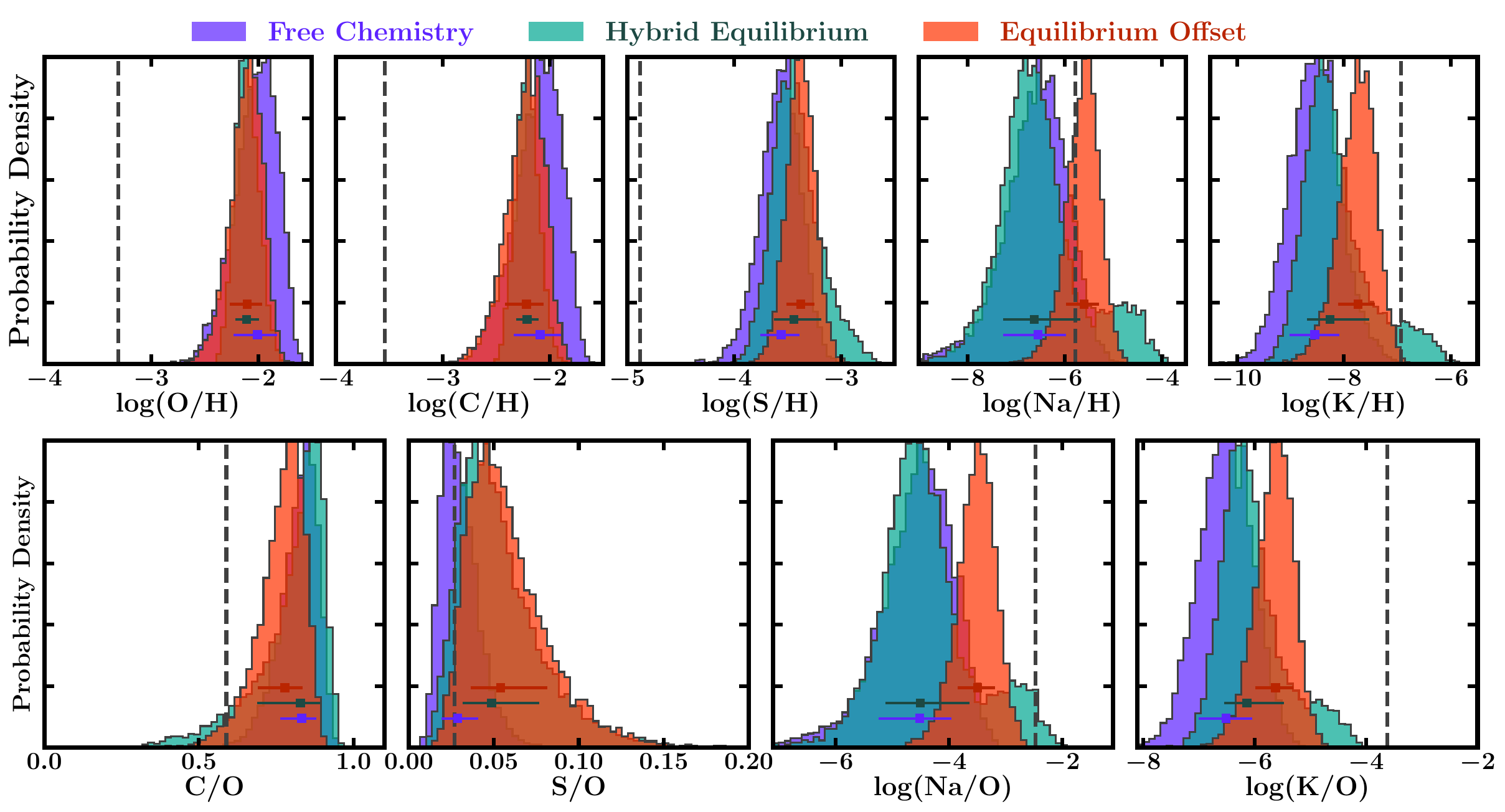}
    \caption{Elemental abundances and ratios of O, C and S, inferred from atmospheric composition constraints obtained with the free chemistry, hybrid equilibrium and equilibrium offset retrievals presented in sections \ref{vira_subsec:free_chemistry_retrievals}, \ref{vira_subsec:hybrid_equilibrium} and \ref{vira_subsec:equilibrium_offset_rets}. The horizontal errorbars denote the median and corresponding 1-$\sigma$ interval, while the vertical dashed lines denote the corresponding solar value. The above distributions do not consider any oxygen that may be sequestered in refractory silicates.}
    \label{vira_fig:elemental_abundance_distributions}
\end{figure*}

With our hybrid equilibrium chemistry retrieval, we obtain constraints for $[\frac{\mathrm{O}}{\mathrm{H}}]$, $[\frac{\mathrm{C}}{\mathrm{H}}]$ and $[\frac{\mathrm{S}}{\mathrm{H}}]$ as well as the log-mixing ratios of SO$_2$, Na and K. Specifically, our retrieval constrains $ [\frac{\mathrm{O}}{\mathrm{H}}] $, i.e. oxygen abundance relative to solar in dex, to $+1.21^{+0.12}_{-0.10}$ and $[\frac{\mathrm{C}}{\mathrm{H}}]$ to $+1.33^{+0.11}_{-0.10}$. Together, the two constraints at face value indicate a C/O ratio of $0.83^{+0.06}_{-0.14}$, which is higher than the solar value of 0.59 \citep{Asplund2021}. Our retrieval also constrains $[\frac{\mathrm{S}}{\mathrm{H}}]$ to $+1.47^{+0.24}_{-0.18}$. This is higher than the values retrieved for the relative enhancements of O and C, with the retrieved S enhancement estimate lying within 1-$\sigma$ of that retrieved for C, and beyond 1-$\sigma$ of that of O. The absolute retrieved elemental abundance constraints correspond to an S/O ratio of $0.049^{+0.028}_{-0.017}$. Compared to the equivalent value for solar elemental abundances, at 0.0269 for S/O \citep{Asplund2021}, our results at this stage taken at face value indicates a super-solar S/O ratio. We emphasise that these constraints are not definitive and will be further refined in subsequent retrievals. Meanwhile, the retrieval does not obtain any meaningful constraints for $[\frac{\mathrm{N}}{\mathrm{H}}]$.

Beyond elemental abundances, our retrieval constrains the log-mixing ratio of SO$_2$ to $-5.49^{+0.18}_{-0.16}$. Na and K meanwhile are constrained to log-mixing ratios of $-6.40^{+0.93}_{-0.64}$ and $-8.04^{+0.73}_{-0.43}$. All four estimates are within 1-$\sigma$ of those obtained with our earlier free chemistry retrievals.

Based on our retrieval results, we construct the retrieved vertical mixing ratio profiles for the 7 molecules whose abundances are set by the chemical equilibrium calculation, as described in \ref{vira_subsubsec:constraint_postprocessing}. The abundance profiles for molecules with significant opacity contributions to the final spectrum are shown in Figure \ref{vira_fig:retrieved_Xprofiles_PT}. We find that the retrieved profiles are largely consistent with the constant-with-altitude free chemistry constraint within some regions, while also significantly deviating in others.

We note that unlike in the free retrieval case, we do not find any significant amount of HCN being present in our retrieved vertical mixing ratio profiles. This is due to the equilibrium chemistry calculation, which gives rise to significant HCN abundances only for very high C/O ratios and higher temperatures. Given the C/O ratio is dictated by the very prominent H$_2$O, CO and CO$_2$ features, the retrieval is unable to invoke an HCN mixing ratio high enough to contribute to the spectral fit.

Similarly to the free chemistry retrievals, we once again obtain constraints for the mixing ratio of ZnS aerosols. Specifically, we retrieved a log-mixing ratio of $-7.99^{+0.66}_{-0.49}$, with a corresponding modal particle radius $\mathrm{log}(r_\mathrm{c} / \mu \mathrm{m}) = -1.88^{+0.09}_{-0.12}$, effective scale factor $H_\mathrm{c} = 0.61^{+0.04}_{-0.03}$ and terminator coverage fraction $f_\mathrm{c} = 0.68^{+0.05}_{-0.05}$. These constraints are consistent with those obtained in the free chemistry retrievals. They are however more precise, particularly in the case of the modal particle radius.

Our retrieved P-T profile is somewhat hotter than that obtained in the free retrieval case, as shown in Figure \ref{vira_fig:retrieved_Xprofiles_PT}, with a 1-$\sigma$ temperature range of $\sim760-860$~K at the top of the  observed photosphere (2-$\sigma$ pressure range), rising to $\sim990-1050$~K at the bottom. The P-T profile is not consistent with an isotherm, in this case showing a steeper rise within the observed photospheric pressure range.

As evidenced by our earlier free retrievals, which do not produce any constraints for the mixing ratio of CH$_4$, the spectrum does not display CH$_4$ absorption features to any significant extent. This is a primary cause for the present hybrid equilibrium retrieval invoking a higher atmospheric temperature than in the free chemistry cases. This is done so that the abundance of CH$_4$ in the model is reduced, as CH$_4$ reacts with H$_2$O to form CO at higher temperatures, in order to match the data.

We therefore find that switching to a more refined atmospheric model, which considers the mixing ratio profiles of prominent molecules, can lead to a more nuanced understanding of the atmospheric properties. Our findings also highlight the strong coupling between the atmospheric composition and the temperature profile. As noted in section \ref{vira_subsubsec:hybrid_equilibrium_chemistry}, the assumption of chemical equilibrium can also introduce biases, especially in cases where there are significant non-equilibrium processes at play. Moreover, this approach does not indicate which species are the main drivers of the retrieved elemental abundances. As such, the atmospheric properties retrieved with our hybrid equilibrium retrievals are not definitive. Nevertheless, the retrieved atmospheric constraints provide important insights into the atmospheric properties of WASP-39~b, which can also serve as starting points for the following stage in the retrieval cascade, which does not suffer from these biases.

\subsection{Equilibrium Offset Retrievals}
\label{vira_subsec:equilibrium_offset_rets}

Equilibrium offset retrievals, described in section \ref{vira_subsubsec:offset_retrievals}, combine the flexibility of free chemistry retrievals and the sophisticated vertical mixing ratio profiles of hybrid equilibrium retrievals. Following our cascading paradigm, the present retrievals continue building on the constraints obtained with prior retrievals. We set the fiducial elemental abundance of O, C and S to the median values obtained with our hybrid equilibrium retrieval, at $[\frac{\mathrm{O}}{\mathrm{H}}]$ = +1.2, $[\frac{\mathrm{C}}{\mathrm{H}}]$ = +1.3 and $[\frac{\mathrm{C}}{\mathrm{H}}]$ = +1.5. Additionally, as $[\frac{\mathrm{N}}{\mathrm{H}}]$ was unconstrained, we nominally set it to a value of +1, i.e. roughly the same degree of enrichment as O and C.   

The retrieved atmospheric model relies on an equilibrium chemistry calculation, in order to account for the variation in vertical mixing ratio profiles with temperature. As such, it can be expected that the coupling between the atmospheric composition and the temperature profile seen before will persist, even though the model is significantly more flexible than in the hybrid equilibrium retrieval case. In order to ensure that our obtained retrieved results are maximally robust, we carry out retrievals using both the MultiNest and UltraNest retrieval packages.

The retrieved spectral fit obtained with MultiNest is shown in Figure \ref{vira_fig:eq_offset_spectrum_photosphere}. With this MultiNest retrieval, we obtain precise constraints for the equilibrium offsets for H$_2$O and CO, at $\mathrm{log(}\delta_{\mathrm{H}_2\mathrm{O}}\mathrm{)} = 0.04^{+0.14}_{-0.14}$ and $\mathrm{log(}\delta_{\mathrm{CO}}\mathrm{)} = -0.05^{+0.17}_{-0.21}$. Both are consistent with no offset, indicating that they are the main drivers of the retrieved elemental abundances in our earlier hybrid equilibrium retrieval. The retrieval additionally constrains the multiplicative offset for the  CO$_2$ vertical mixing ratio profile to $-0.15^{+0.16}_{-0.16}$. As such, while CO$_2$ may be slightly depleted, it is nevertheless consistent with thermochemical equilibrium expectations. Lastly, H$_2$S is retrieved with a multiplicative offset of $0.03^{+0.13}_{-0.14}$, which is also consistent with no offset. 

A notable difference from the hybrid equilibrium retrieval is the retrieved P-T profile. For the present retrieval, the temperature of the top of the  photosphere (2-$\sigma$) is constrained to a 690-730~K 1-$\sigma$  temperature range, while the 1-$\sigma$ temperature range at the bottom of the photosphere ranges is $\sim$740-790~K, as can be seen in Figure \ref{vira_fig:retrieved_Xprofiles_PT}. The temperature profile is therefore cooler than that retrieved with the hybrid equilibrium retrieval, and largely consistent with that retrieved in the free chemistry retrieval.

The retrieved temperature profile consists of temperatures at which CH$_4$ is expected to be a primary carbon carrier. As mentioned above, the hybrid equilibrium retrieval invoked a higher temperature in order to reduce the mixing ratio of CH$_4$ to match the observations, which do not contain any absorption features from it. The present equilibrium offset retrieval however can instead invoke the temperature profile that best explains the data, while reducing CH$_4$ through its equilibrium offset. This is borne out of the retrieved posterior distribution for $\mathrm{log(}\delta_{\mathrm{CH}_4}\mathrm{)}$, which extends solely over values corresponding to significant depletions. Specifically, we find a minimum depletion factor of $2.7\times 10^{-4}$ at 99\% confidence, i.e. CH$_4$ is depleted by at least $\sim$3.6 dex from equilibrium expectation values.

The retrieved temperature profile is also close to temperatures at which a non-negligible part of the atmospheric nitrogen budget goes towards NH$_3$ rather than solely N$_2$. Similarly to CH$_4$, however, the retrieval only produces an upper bound for $\mathrm{log(}\delta_{\mathrm{NH}_3}\mathrm{)}$, with a 99\% confidence minimum depletion factor of 0.74, i.e. 0.13 dex below equilibrium expectations. This is significantly higher than the limit obtained for CH$_4$, due to the lower starting point for the NH$_3$ mixing ratio profile, as well as NH$_3$ being less spectrally prominent and therefore requiring a higher mixing ratio before it contributes observable spectral features. Meanwhile, the equilibrium offset parameters for HCN and C$_2$H$_2$ remain unconstrained.

For SO$_2$, Na and K, which were included in the model as free chemistry parameters, we retrieve log-mixing ratio values of $-5.73^{+0.19}_{-0.22}$, $-5.40^{+0.32}_{-0.38}$ and $-7.51^{+0.33}_{-0.39}$. All three estimates are consistent with those obtained with the hybrid equilibrium retrieval above. However, in the case of Na and K, this is largely due to the significantly larger 1-$\sigma$ errors obtained in the hybrid equilibrium retrieval, while the present retrieval obtains significantly more precise constraints at $\sim$0.3-0.4~dex. As such, even though Na and K were included as free chemistry parameters in both cases, the treatment of the other chemical species in the atmosphere affects their retrieved mixing ratio constraints.

Having obtained precise and detailed constraints for the vertical composition profile of the atmosphere, we construct the median retrieved atmospheric photosphere, using the approach detail in \ref{vira_subsubsec:constraint_postprocessing}. We find that the pressures primarily giving rise to the observed transmission spectrum lie between $\sim 10^{-2}-10^{-3}$~bar, as shown in Figure \ref{vira_fig:eq_offset_spectrum_photosphere}. Notably, the peak and wings of the prominent CO$_2$ absorption feature at $\sim$4.5~$\mu$m together probe a pressure range of more than 2~dex. As such, this feature is highly informative in constraining the vertical variation of the CO$_2$ abundance as well as the temperature profile. Additionally, the SO$_2$ feature at $\sim$4~$\mu$m probes a pressure of $\sim$10~mbar. As SO$_2$ is expected to be produced primarily due to photochemistry, this implies that photochemical effects remain significant down to altitudes corresponding to $\sim$10~mbar.

Given the complex nature of the atmospheric model used, we verify our findings by carrying out an additional retrieval using the UltraNest nested sampling algorithm as the final step in our retrieval cascade, as described in section \ref{vira_subsec:ultranest}. We obtain chemical constraints that are in good agreement with those obtained with MultiNest. For H$_2$O, CO, CO$_2$ and H$_2$S, we find that $\mathrm{log(}\delta_{\mathrm{H}_2\mathrm{O}}\mathrm{)} = -0.02^{+0.11}_{-0.13}$, $\mathrm{log(}\delta_{\mathrm{CO}}\mathrm{)} = -0.07^{+0.17}_{-0.19}$, $\mathrm{log(}\delta_{\mathrm{CO_2}}\mathrm{)} = -0.27^{+0.18}_{-0.18}$ and $\mathrm{log(}\delta_{\mathrm{H_2S})} = -0.06^{+0.17}_{-0.19}$. Additionally, CH$_4$ is once again found to be significantly depleted relative to thermochemical equilibrium expectations, with a 99\% confidence minimum depletion of 4.0~dex.

Beyond the atmospheric composition, the UltraNest retrieval constrains the same Mie scattering aerosol and temperature profile properties as those obtained with MultiNest. UltraNest however also identifies a second mode in the posterior distribution, corresponding to a lower ZnS log-mixing ratio of $\sim$-16~dex and a modal particle radius of $\sim$0.1~$\mu$m, along with a log-mixing ratio of $\sim$-20~dex. This also corresponds to a temperature profile that rises more steeply from a temperature of $\sim$600~K at the top of the atmosphere up to $\sim$~1100~K at the bottom. We note that the retrieved chemical constraints detailed above are unaffected if we isolate either of the two aerosol and temperature modes UltraNest obtains.

\subsection{Elemental Composition and Chemical Disequilibrium}
\label{vira_subsec:atmosphere_of_wasp39b}

\begin{figure}
    \centering
    \includegraphics[angle=0,width=0.48\textwidth]{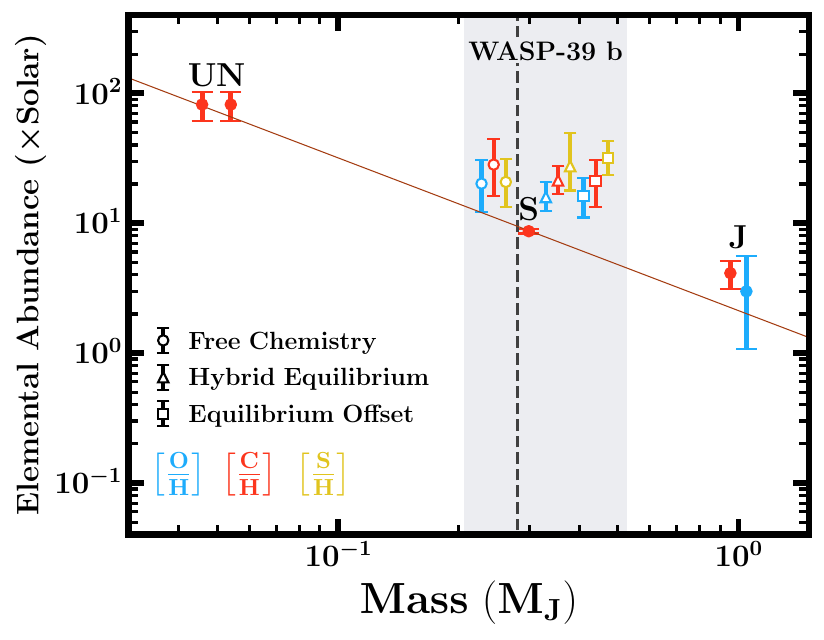}
    \caption{The inferred O, C and S elemental abundance ratios for WASP-39~b obtained with the free chemistry, hybrid equilibrium and equilibrium offset retrievals, all using the canonical Mie scattering aerosol species. Also shown is the mass-metallicity trend for solar system giant planets derived from their CH$_4$ measurements \citep{Atreya2022}, as well as Jupiter's oxygen abundance derived from the H$_2$O measurement of Juno \citep{Li2020}.  WASP-39~b has a mass of 0.28~M$_\mathrm{J}$, as indicated by the vertical dotted line. The errorbars showing elemental abundance ratios for WASP-39~b are offset from this value for visual clarity.}
    \label{vira_fig:mass_metallicity}
\end{figure}

Having run the full extent of our atmospheric retrieval cascade, we are now able to combine our findings to obtain a detailed picture of the atmosphere of WASP-39~b. In particular, we use the retrieved atmospheric composition from all stages to infer the elemental abundances they correspond to. Additionally, we use our retrieved constraints to establish how significant disequilibrium effects are in shaping the atmospheric composition. We note that all our findings pertain to the observed photosphere, which ranges between pressures of $\sim10^{-2}-10^{-4.5}$~bar, as shown in Figure \ref{vira_fig:eq_offset_spectrum_photosphere}.

\begin{table*}
    
    \centering
    \caption{Elemental abundances and ratios obtained from the retrieval cascade of WASP-39~b presented in section \ref{vira_sec:WASP-39b_retrievals}, inferred from averaged chemical abundances across the observed photosphere. The solar system elemental abundances and ratios are derived from \citet{Asplund2021}.}
    \begin{tabular}{c|c|c|c|c|c|c|c|c|c}
          & log(O/H) & log(C/H) & log(S/H) & log(Na/H) & log(K/H) & C/O & S/O & log(Na/O) & log(K/O)\\
         \hline
         \hline
        \multicolumn{10}{c}{{\it Free Chemistry}} \\
        \hline

        Sigmoid Clouds & $-2.04^{+0.13}_{-0.15}$ & $-2.22^{+0.16}_{-0.19}$ & $-3.75^{+0.19}_{-0.21}$ & $-6.72^{+0.64}_{-1.87}$ &  $-8.55^{+0.46}_{-0.48}$ & $0.66^{+0.09}_{-0.11}$ & $0.020^{+0.012}_{-0.008}$ & $-4.67^{+0.61}_{-1.86}$ & $-6.5^{+0.44}_{-0.45}$ \\[0.2cm]
        Mie Aerosols & $-2.01^{+0.18}_{-0.22}$ & $-2.09^{+0.20}_{-0.24}$  &  $-3.56^{+0.18}_{-0.19}$ & $-6.55^{+0.58}_{-0.72}$ & $-8.55^{+0.46}_{-0.48}$ &  $0.83^{+0.05}_{-0.07}$ & $0.029^{+0.012}_{-0.009}$ & $-4.51^{+0.56}_{-0.73}$ & $-6.51^{+0.47}_{-0.49}$  \\
        \hline 
        \multicolumn{10}{c}{{\it Hybrid Equilibrium}} \\
        \hline
        Mie Aerosols & $-2.11^{+0.12}_{-0.10}$ & $-2.21^{+0.11}_{-0.10}$ & $-3.44^{+0.25}_{-0.19}$ & $-6.63^{+0.95}_{-0.64}$ & $-8.26^{+0.73}_{-0.43}$ & $0.83^{+0.06}_{-0.14}$ & $0.049^{+0.028}_{-0.017}$ & $-4.50^{+0.87}_{-0.61}$ & $-6.14^{+0.66}_{-0.41}$\\
        \hline
        \multicolumn{10}{c}{{\it Equilibrium Offset}} \\
        \hline
        MultiNest & $-2.10^{+0.14}_{-0.17}$ & $-2.22^{+0.16}_{-0.20}$  & $-3.38^{+0.13}_{-0.14}$ & $-5.60^{+0.31}_{-0.37}$ & $-7.74^{+0.32}_{-0.37}$ & $0.78^{+0.06}_{-0.09}$ & $0.054^{+0.028}_{-0.018}$ & $-3.49^{+0.31}_{-0.35}$ & $-5.62^{+0.31}_{-0.37}$\\ [0.2cm]
        UltraNest & $-2.17^{+0.15}_{-0.16}$ & $-2.29^{+0.17}_{-0.19}$ & $-3.47^{+0.17}_{-0.19}$& $-5.88^{+0.51}_{-0.64}$ & $-7.91^{+0.41}_{-0.47}$ & $0.76^{+0.06}_{-0.07}$ & $0.050^{+0.023}_{-0.016}$ &$-3.73^{+0.46}_{-0.57}$ &$-5.72^{+0.34}_{-0.45}$\\ 
        \hline
        \hline
        \multicolumn{10}{c}{{\it Solar Values}} \\
        \hline
        &-3.31 & -3.54 & -4.88 & -5.78 & -6.93 & 0.59 & 0.0269 & -2.47 & -3.62 \\
        \hline
    \end{tabular}
    
    \label{tab:inferred_elemental_ratios}
\end{table*}

\subsubsection{Atmospheric Elemental Inventory}
\label{vira_subsubsec:atmospheric_elemental_inventory}

The retrievals carried out in this work obtain precise abundance constraints for a range of chemical species, including those expected to be primary O, C and S-bearing molecules under thermochemical equilibrium \citep[e.g.][]{Burrows&Sharp1999, Lodders&Fegley2002, Zahnle2009}. As such we are able to infer the corresponding elemental abundances and ratios. In the case of the hybrid equilibrium and equilibrium offset retrievals, we infer the atmospheric elemental inventory by averaging across the observed photosphere.

The inferred elemental abundances and ratios of O, C and S obtained with the free chemistry, hybrid equilibrium and equilibrium offset retrievals are presented in Table \ref{tab:inferred_elemental_ratios}. The probability distributions corresponding to these constraints are also shown in Figure \ref{vira_fig:elemental_abundance_distributions} along with those of Na and K, which are less robustly constrained across our retrieval cascade. With the free chemistry and Mie scattering retrieval, we retrieve atmospheric composition constraints with log(O/H) = $-2.01^{+0.18}_{-0.22}$, log(C/H) = $-2.09^{+0.20}_{-0.24}$, log(S/H) = $-3.56^{+0.18}_{-0.19}$, log(Na/H) = $-6.55^{+0.58}_{-0.72}$ and log(K/H) = $-8.55^{+0.46}_{-0.48}$. Meanwhile, the elemental abundances inferred from the offset retrieval carried out using UltraNest are log(O/H)=$-2.17^{+0.15}_{-0.16}$, log(C/H) = $-2.29^{+0.17}_{-0.19}$, log(S/H) = $-3.47^{+0.17}_{-0.19}$, log(Na/H) = $-5.88^{+0.51}_{-0.64}$ and log(K/H) = $-7.91^{+0.41}_{-0.47}$. We note that the abundances of Na and/or K are less robust relative to the other species, as they are based on a single absorption line feature and can be sensitive to the cloud parametrisation.

Additionally, all retrieval configurations obtain O, C and S abundances that are enhanced by more than 1~dex from solar values. The free chemistry and Mie scattering retrieval specifically obtains O/H, C/H and S/H abundances corresponding to $20.14^{+10.54}_{-8.08}$, $28.22^{+16.32}_{-12.12}$ and $20.76^{+10.3}_{-7.47}$ $\times$ solar, respectively. Meanwhile, from the UltraNest equilibrium offset retrieval we infer elemental enhancements of $13.93^{+5.66}_{-4.3}$, $17.78^{+8.58}_{-6.23}$ and $25.92^{+12.69}_{-9.25}$ $\times$ solar, for O/H, C/H and S/H, respectively. The retrieved Na abundances are slightly sub-solar to solar, with the free chemistry and Mie scattering retrieval obtaining an Na/H abundance that is $0.171^{+0.476}_{-0.139} \times$ solar, and the UltraNest equilibrium offset retrieval inferring a $0.786^{+1.734}_{-0.607} \times$ solar Na abundance. Lastly, all retrievals obtain a sub-solar K/H abundance, corresponding to $0.0242^{+0.0461}_{-0.0161} \times$ solar in the case of the free chemistry and Mie scattering retrieval, and $0.105^{+0.167}_{-0.069} \times$ for the UltraNest equilibrium offset retrieval.

The inferred atmospheric C/O and S/O ratios are marginally super-solar across the retrievals. With the retrieval using free chemistry and Mie scattering aerosols, we find C/O and S/O ratios of $0.83^{+0.05}_{-0.07}$ and $0.029^{+0.012}_{-0.009}$, while the UltraNest equilibrium offset retrieval obtains ratios of $0.76^{+0.06}_{-0.07}$ and $0.050^{+0.023}_{-0.016}$ for C/O and S/O, respectively. Additionally, the retrieved Na/O and K/O abundances are sub-solar across all retrievals. We note that the present constraints do not account for any oxygen Na, K or other elements that may be sequestered in refractories such as silicates or other compounds.

Prior analyses of various single-transit JWST observations without using atmospheric retrievals also found that the atmosphere of WASP-39~b has an overall super-solar metallicity \citep{Ahrer2023, Alderson2023, Feinstein2023, Rustamkulov2023}. The same forward model analyses of NIRISS, NIRSpec G395H and NIRcam observations found sub-solar to solar C/O ratios \citep{Ahrer2023, Alderson2023, Feinstein2023}, while those for NIRSpec PRISM found the data is best explained by super-solar C/O ratios \citep{Rustamkulov2023}. Our present results are consistent with previous inferences of a super-solar atmospheric metallicity. We note however that care must be exercised when assuming equilibrium chemistry, as disequilibrium abundances may bias the results if not accounted for in the retrievals. Our abundance estimates are also consistent with the prior findings of \citep{Constantinou2023}, obtained with an earlier reduction of a subset of the NIRSpec PRISM data \citep{ERS2023} combined with HST/WFC3. Lastly, we note that \citet{Niraula2023}, using the NIRSpec G395H observations presented by \citet{Alderson2023}, also find a super-solar metallicity, albeit at a significantly higher degree of enrichment compared to the present work.

Figure \ref{vira_fig:mass_metallicity} shows the inferred O, C and S abundances of WASP-39~b obtained with the free chemistry, hybrid equilibrium and equilibrium offset retrievals, relative to those of giant solar system planets. Across all three retrieval cases, the three elemental abundances indicate a metallicity enhancement relative to solar values, and a slightly higher elemental enrichment relative to Saturn's, giving a consistent picture of core accretion in action \citep{Pollack1996}. Notably, the three elemental constraints obtained for WASP-39~b are of comparable precision to those obtained for solar system planets \citep{Atreya2022}, highlighting the potential of transmission spectroscopy in the JWST era.

\subsubsection{Chemical Disequilibrium in the Atmosphere of WASP-39~b}
\label{vira_subsubsec:inference_of_chemical_disequilibrium}

While the temperature across the observed photosphere obtained with the equilibrium offset retrieval ranges between 700-800~K, CO is nevertheless expected to be present under chemical equilibrium, as indicated in Figure \ref{vira_fig:retrieved_Xprofiles_PT}. This is due the relatively low pressure, which favours the production of CO at the expense of CH$_4$. However, at these pressures the timescales over which such reactions take place may be comparable to or longer than the eddy diffusion timescale. As such, a possible explanation for the depletion of both CH$_4$ as well as NH$_3$ is that the observed abundances are not the result of local thermochemical equilibrium, but rather the quenched products of reactions occurring at deeper, hotter atmospheric regions \citep{Prinn1977, Showman2010, Visscher2011}. Deeper down in the atmosphere, CO and N$_2$ dominate over CH$_4$ and NH$_3$, respectively, as their production is favoured due to the much higher temperatures temperatures \citep{ Burrows&Sharp1999, Lodders&Fegley2002, Madhusudhan2011a, Moses2013b}. These species are then lofted up by vertical mixing processes to higher, colder altitudes at a faster rate than they can be removed by equilibrium chemistry \citep{Prinn1977}.

In addition to vertical mixing, the observed depletions may also be the result of photochemistry, with both CH$_4$ and NH$_3$ being susceptible to photodecomposition at high altitudes \citep{Yung1999, Liang2003, Segura2005, GarciaMunoz2007, Moses2011}. Such a scenario would also be consistent with the presence of SO$_2$, which is primarily produced by the photochemical breakup of H$_2$S \citep{Zahnle2009, Wang2017, Hobbs2021, Polman2022, Tsai2023}. However, our retrievals find that H$_2$S is present in significant quantities in the observed photosphere, indicating that the rate at which H$_2$S is broken up is not sufficient to fully deplete it, contrasting with the absence of CH$_4$ and NH$_3$.

\section{Summary and Discussion}
\label{vira_sec:discussion}

We present VIRA, the latest member of the AURA family of retrieval frameworks \citep{Pinhas2018, Welbanks2021, Nixon2022, Constantinou2023}  customised for JWST observations of exoplanet transmission spectra. In addition to inheriting features of prior AURA-family frameworks, VIRA implements new modelling and statistical features in order to make the most of the wealth of information encoded in JWST transmission spectroscopy observations. VIRA is designed to be used in a cascading manner, with several retrievals being carried out in sequence. Earlier steps in the cascade employ more general models, which provide initial insights into the atmospheric composition, temperature structure and other properties. In the case of high quality observations, these early runs inform how the retrieval is to be refined for the next stage of the cascade. As such, the retrieved model remains computationally tractable at each stage of the cascade, despite each retrieval requiring tens of millions of model evaluations.

A key element of the cascading architecture are the three complementary approaches to modelling atmospheric composition implemented in VIRA. A free chemistry approach is used first, to determine which species contribute observable features to the atmospheric spectrum. For high quality data, the atmospheric composition estimates are further refined by considering these species in a retrieval that uses the second approach, hybrid equilibrium. This retrieval calculates the vertical atmospheric composition profile for each forward model by considering its temperature profile and assuming equilibrium chemistry. The retrieved elemental abundances from this stage are then used in the final and most refined retrievals using the equilibrium offset approach, which retrieves the multiplicative offset of each species' mixing ratio profile, calculated based on elemental abundances inferred from previous hybrid equilibrium retrievals. This approach therefore combines the flexibility of the free chemistry approach with the sophistication of the hybrid equilibrium approach.

In addition to gaseous species, VIRA implements two new approaches in modelling the spectral contributions from atmospheric aerosol particles. The first is a modification of the parametric Rayleigh-like hazes and grey clouds model used in HST era retrievals \citep[e.g.][]{Pinhas2018}, to account for opacity variations over large wavelength ranges. This is modelled as a sigmoid modulation to the grey opacity, reducing it at longer wavelengths, with the location and slope of the modulation as free parameters. The second approach is a physical Mie scattering calculation considering both the composition and modal particle size of aerosol particles, as well as their vertical extent and fractional terminator coverage, \citep[e.g.][]{Constantinou2023} . This model can be employed if earlier stages in the cascade with the parametric sigmoid clouds model indicate that there are significant contributions from non-grey aerosol opacity to the observed spectrum, especially if their contributions vary significantly at longer wavelengths. In addition to these two new approaches, VIRA also inherits the parametric Rayleigh-like hazes and grey clouds model of prior AURA-family frameworks.

Beyond model refinements, VIRA also includes a more sophisticated statistical treatment of observations. For observations with significant correlations, such as those obtained with the NIRISS instrument, VIRA can consider the complete covariance matrix of such observations when computing each model's likelihood. Additionally, VIRA can rely on either MultiNest or UltraNest to carry out the nested sampling-based parameter estimation, with the former primarily used during earlier stages of the cascade while the latter is reserved for the final, most complex retrievals to ensure the prior space is robustly explored.

\subsection{Atmospheric Retrievals of WASP-39~b with VIRA}

We demonstrate VIRA's capabilities by analysing the JWST observations of WASP-39~b, obtained with the NIRISS and NIRSpec PRISM instruments \citep{Holmberg2023, Rustamkulov2023}. Given the high precision and large wavelength coverage of the data, we employ all features described in this work, running the full extent of the retrieval cascade. This approach provides significant insights into the atmospheric properties and processes of WASP-39~b:

\begin{itemize}

\item The transmission spectrum of WASP-39~b contains significant spectral contributions arising from H$_2$O, CO$_2$ and SO$_2$, as well as some contributions from CO, H$_2$S, Na and K. These findings are in line with prior results both from JWST observations \citep{Alderson2023, Ahrer2023, Constantinou2023, Feinstein2023, Rustamkulov2023, Niraula2023} as well as HST and ground-based observations \citep{Sing2016, Nikolov2016, Fischer2016, Wakeford2018, Pinhas2019, Kirk2019, Welbanks2019b, Kawashima2021}.

\item Having obtained abundance constraints for the prominent O-, C- and S-bearing molecules, we are able to infer the atmospheric elemental budget. Considering our most conservative molecular constraints, obtained with the free chemistry retrieval, we infer super-solar abundances of log(O/H)=$-2.01^{+0.18}_{-0.22}$, log(C/H)=$-2.09^{+0.2}_{-0.24}$ and log(S/H)=$-3.56^{+0.18}_{-0.19}$. Specifically, we find that the inferred elemental abundances represent enhancements of $20.1^{+10.5}_{-8.1}\times$, $28.2^{+16.3}_{-12.1}\times$ and $20.8^{+10.3}_{-7.5}\times$ solar elemental abundances, for O/H, C/H and S/H. Moreover, the present constraints correspond to C/O and S/O ratios of $0.83^{+0.05}_{-0.07}$ and $0.029^{+0.012}_{-0.009}$, which are largely consistent with super-solar values \citep[C/O = 0.59, S/O = 0.027;][]{Asplund2021}. We note however that some of the oxygen budget may be sequestered in refractory silicates, in which case the C/O ratio may be consistent with solar values.

\item Our retrievals suggest significant spectral contributions from ZnS aerosols in the spectrum of WASP-39~b. This is consistent with prior works using both simulated and actual JWST spectra finding that their precision and wide wavelength coverage can in principle lead to constraints on the aerosol properties \citep{Benneke2013, Wakeford2015, Pinhas2017, Mai2019, Lacy2020a, Constantinou2023}. We note that while the retrievals carried out in \citet{Constantinou2023} invoke MgSiO$_3$ aerosols over ZnS, they were carried out on a much more spectrally limited dataset, using HST/WFC3 instead of NIRISS. Additionally, the prior indications of MgSiO$_3$ aerosols were found to be dependent on the specific choice of reduction pipeline.

\item CH$_4$ is found to be significantly depleted relative to its expected mixing ratio under chemical equilibrium. We specifically retrieve a depletion of least 3.8~dex relative to equilibrium chemistry expectations, at a 99\% limit. This is in agreement with prior studies determining an absence of CH$_4$ absorption features in its atmosphere \citep{ERS2023, Ahrer2023, Alderson2023, Feinstein2023, Rustamkulov2023}. The significant retrieved depletion of CH$_4$ is likely indicative of disequilibrium effects, as discussed in section \ref{vira_subsubsec:inference_of_chemical_disequilibrium}.

\item We find that the effective photosphere giving rise to the observed transmission spectrum ranges between pressures of $\sim10^{-1.8}-10^{-4.4}$~bar, as shown in Figure \ref{vira_fig:eq_offset_spectrum_photosphere}. Prominent spectral features, such as that of CO$_2$ at $\sim4.5$~$\mu$m therefore probe a pressure range of $\sim$2.5~dex. Additionally, the $\sim$4~$\mu$m absorption feature of SO$_2$, which is formed through photochemistry \citep{Zahnle2009, Tsai2023}  corresponds to a minimum pressure of $\sim10^{-2}$~bar, indicating that photochemical processes are significant across a significant part of the observed photosphere.

\item The retrieved non-isothermal temperature profile has a 1-$\sigma$ temperature range within the photosphere of $\sim$700-800~K for the free chemistry retrieval, $\sim$750-1100~K for the hybrid equilibrium retrieval and $\sim$700-800~K for the equilibrium offset retrieval. We find that assuming equilibrium chemistry imposes strong coupling between the temperature profile and the atmospheric composition, particularly in the presence of significant disequilibrium effects. In the present case, the hybrid equilibrium chemistry retrieval was driven to higher temperatures than other retrievals in order to reduce the CH$_4$ abundance to match the observations.

\item We find that the NIRISS data spanning 0.6-2.8$\mu$m primarily contain information on the abundances of H$_2$O, Na and K as well as the macro- and microscopic properties of atmospheric aerosols. Meanwhile the NIRSpec PRISM observations, which in the present case span 2-5.5$\mu$m, are primarily informative for the abundances of CO, CO$_2$, H$_2$S, SO$_2$. Both datasets together are sensitive to non-uniform vertical mixing ratio profiles and non-isothermal P-T profiles.

\item  Finally, VIRA incorporates the capability for robust Bayesian inference using nested sampling in order to fully leverage a highly sophisticated atmospheric model. In the present case study of WASP-39~b, we find that while MultiNest and UltraNest both obtain consistent atmospheric composition constraints, UltraNest offers more robust estimates of the final posterior distribution, and is valuable for robustness checks in the case of extensive high-quality observations.

\end{itemize}

\subsection{On Multi-Elemental Constraints in the JWST Era}

The present work demonstrates how broadband JWST transmission spectroscopy observations of irradiated gas giant exoplanets can yield simultaneous constraints for a range of elemental species, in this case O, C and S, as well as tentative constraints for Na and K. Moreover, as shown in Figure \ref{vira_fig:mass_metallicity}, the inferred abundance constraints for these elements are of comparable precisions to those of C abundance measurements of several solar system gas giant planets \citep{Atreya2022}, and more precise than the Juno measurement of O in Jupiter \citep{Li2020}. Our work therefore demonstrates the capability of JWST transmission spectroscopy to yield simultaneous abundance constraints for a broad range of elemental species, which are challenging to obtain for solar system giant planets. Such constraints could enable important insights into the formation and migration mechanisms of giant exoplanets \citep{Madhusudhan2019}, as well as provide a comparative benchmark for solar system planets.

\subsection{Expanding the Retrieval Cascade}

The retrieval cascade considered in this work focuses on refining the treatment of atmospheric chemistry and aerosol scattering in the model. As noted above, this was motivated by both the precision and wavelength coverage of JWST observations of WASP-39~b, which enable such sophisticated constraints. Moreover, WASP-39~b is close to the transitional temperature where either CO or CH$_4$ are dominant under thermochemical equilibrium, requiring careful consideration of equilibrium and disequilibrium effects in characterising its atmosphere.

We emphasise that the cascading retrieval architecture presented in this work can be readily expanded beyond atmospheric chemistry and aerosols on a case-by-case basis  using AURA-family functionality \citep{Pinhas2018, Welbanks2021, Nixon2022, Constantinou2023}. As noted in Section \ref{vira_subsec:cascading_architecture}, other refinements can include 3D effects, which can be significant for very strongly irradiated gas giants, or stellar heterogeneities, which can affect transmission spectroscopy of planets around active M dwarfs. The cascading approach means that such considerations are included in the model in a robust way, assessing how their inclusion affects other retrieved atmospheric constraints. Moreover, the cascading architecture means future advancements in atmospheric modelling can also be incorporated in VIRA while efficiently managing this increase in sophistication and complexity.

\subsection{Caveats and Further Considerations}

Qualitatively, vertical mixing and photochemistry affect the vertical atmospheric composition profiles in different ways. Vertical mixing brings up species that are present at higher pressures and temperatures, and in extreme cases can effectively result in mixing ratio profiles that are largely constant above the quenching pressure. Photochemistry meanwhile is expected to modify the atmospheric composition profiles at higher altitudes, depleting some species and enriching others. In reality, both processes are likely to be at play, potentially resulting in significant variation in the vertical mixing ratios of several species. The equilibrium offset approach implemented in VIRA treats equilibrium chemistry abundance profiles as starting points, with the offsets themselves effectively being zeroth order corrections. This approach is therefore expected to be accurate in the case of small deviations from thermochemical equilibrium, as well as for significant depletions, like that of CH$_4$ in the present work, where the abundance profile's shape does not matter. This approach however may not be accurate for intermediate deviations from thermochemical equilibrium, which may result in significantly different abundance profiles. 

For even more precise observations than those considered in the present work, even more refined atmospheric abundance models may be required. In the limiting case of extremely precise spectra, co-adding multiple transits, the suitability of the present approaches and any further refinements can be benchmarked against simulated observations generated from global circulation models and/or detailed chemical disequilibrium models.

Our effective photosphere calculations presented in section \ref{vira_subsec:hybrid_equilibrium} highlight the extensive range of pressures probed when retrieving on observations spanning a large wavelength range. In the case of WASP-39~b specifically, troughs between absorption peaks can probe pressures as high as $\sim$10$^{-1.8}$~bar, especially in regions without significant Mie scattering aerosol opacity. Meanwhile, large spectral features, particularly that of CO$_2$ at $\sim$4.5~$\mu$m, can probe pressures as low as $\sim 10^{-5}$~bar. It is expected that transmission spectra of more enriched atmospheres, such as those of sub-Neptune planets, will probe lower pressures, with the opposite being true for lower metallicity atmospheres. 

Our work demonstrates the significant increase in sophistication demanded of modern retrieval frameworks, if they are to rise to the challenge of making the most of JWST observations. Moreover, it highlights the immense amount of information that can be encoded in a JWST spectrum of an exoplanet atmosphere which when analysed can lead to significant insights into the atmospheric properties of exoplanets. The modelling and statistical features implemented in the VIRA retrieval framework and presented in this work were directly motivated by the first JWST observations, which are of a hot Saturn. As JWST proceeds to observe a wide variety of planets, from Ultra-hot Jupiters to temperate sub-Neptunes, retrieval frameworks must be capable of adapting and innovating in the face of unprecedented observations. As such, the features presented here will inevitably be expanded upon in the future, as we proceed in the golden era of JWST.

\section*{Acknowledgements}

We thank the anonymous reviewer for their valuable comments on the manuscript. This work was performed using resources provided by the Cambridge Service for Data Driven Discovery (CSD3) operated by the University of Cambridge Research Computing Service (\url{www.csd3.cam.ac.uk}), provided by Dell EMC and Intel using Tier-2 funding from the Engineering and Physical Sciences Research Council (capital grant EP/P020259/1), and DiRAC funding from the Science and Technology Facilities Council (\url{www.dirac.ac.uk}).

\section*{Data Availability}
This work is based on observations made with the NASA/ESA/CSA JWST. No new data were generated in support of this research.



\bibliographystyle{mnras}

\bibliography{refs} 



\appendix

\section{A Retrieval Cascade on Simulated Observations}
\label{vira_sec:appendix}

We present a retrieval cascade on simulated observations, demonstrating how the cascading architecture leads to a model capable of obtaining accurate atmospheric constraints. 

Table \ref{tab:simulated_retrieval_evidences} shows the Bayesian evidence obtained at each stage of the retrieval cascade. While the Bayesian evidence is a good indicator of the spectral fit, which we intend to improve as we descend the retrieval cascade, we emphasise that it does not fully capture the physical plausibility of the models considered. While the hybrid equilibrium and equilibrium offset retrievals provide comparable evidence, the latter is more flexible and representative.

\begin{table}
    
    \centering
    \caption{ Retrieved Bayesian evidence at each stage of the retrieval cascade on simulated observations.}
    \begin{tabular}{c|c}
         Model & log-Bayesian Evidence \\
         \hline
         \hline
        \multicolumn{2}{c}{{\it Free Chemistry}} \\
        \hline

        Grey Clouds & $1702.1 \pm 0.2$\\[0.2cm]
        Sigmoid Clouds & $1713.7 \pm 0.2$\\[0.2cm]
        Mie Aerosols &  $1719.6 \pm 0.2$\\[0.2cm]
        \hline 
        \multicolumn{2}{c}{{\it Hybrid Equilibrium}} \\
        \hline
        Mie Aerosols &  $1732.2 \pm 0.2$\\
        \hline
        \multicolumn{2}{c}{{\it Equilibrium Offset}} \\
        \hline
        MultiNest & $1728.4 \pm 0.2$\\ [0.2cm]
        UltraNest & $1729.8 \pm 0.5$\\ 

        \hline
    \end{tabular}
    
    \label{tab:simulated_retrieval_evidences}
\end{table}

\subsection{Simulated Observations}

We use the same observing configuration as used in the main body of this work, comprising of JWST NIRISS between $\sim$0.6- and NIRSpec PRISM. We use the same uncertainties and covariance matrix as the observations obtained for WASP-39~b and presented by \citet{Holmberg2023} for NIRISS and \citet{Rustamkulov2023} for NIRSpec PRISM. To generate the simulated observations, we compute a synthetic transmission spectrum using the forward model generator of VIRA, which is then binned down to match the existing observations. In both cases, noise is added by sampling a multivariate Gaussian probability distribution, using the covariance matrix presented by \citet{Holmberg2023} for NIRISS and a diagonal covariance matrix for NIRSpec PRISM.

We  consider an H$_2$-rich atmosphere which additionally consists of H$_2$O, CH$_4$, CO, CO$_2$, NH$_3$, H$_2$S, SO$_2$, HCN, C$_2$H$_2$ as well as Na and K. Using the equilibrium offset composition framework, we assume a 10$\times$~solar elemental abundance for O, C, N and S, while enhancing the abundance of CO by a factor of 50 above equilibrium values and depleting that of CH$_4$ by a factor of 100. Such a scenario corresponds to vertical mixing bringing up CO from deeper and hotter regions of the atmosphere where it is abundant, while also resulting in a depletion of CH$_4$. CH$_4$ can additionally be further depleted by photochemical processes at high altitudes.

We consider an atmospheric temperature structure that is non-isothermal, starting at 550~K at the top of the atmosphere and rising monotonically to a temperature of $\sim$950~K at 1 bar. As such, the thermal structure is comparable to that retrieved for WASP-39~b using actual observations, as shown in Figure \ref{vira_fig:retrieved_Xprofiles_PT}, but with a steeper increase in temperature for demonstration purposes.

Similarly to the actual atmospheric constraints WASP-39~b, we consider a partially cloudy atmosphere. We specifically consider a 50\% fractional coverage of ZnS aerosols, with a full vertical extent, a modal particle size of 2 $\mu$m and a mixing ratio of $10^{-8}$.

\subsection{A Canonical Set of Retrieved Gaseous Species}

We begin the cascade by considering the simplest and most general models. We use the free chemistry approach to atmospheric composition and parametric grey clouds and Rayleigh-like hazes. We begin with a large number of chemical species before narrowing down to a canonical set, depending on whether their mixing ratios are reasonably constrained or if they're expected to be present based on theoretical expectations. For this specific case, we obtain constraints for H$_2$O, CH$_4$, CO, CO$_2$, SO$_2$, Na and K, which are all included in the canonical set of retrieved species. We additionally retain HCN and C$_2$H$_2$, which are key markers of the atmospheric C/O ratio and any upper abundance estimates may still be informative, as well as NH$_3$ which can be the dominant N-carrying molecule. Lastly, while the retrieval does not constrain the abundance of H$_2$S, we opt to include it in the atmospheric model as it is expected to be present as indicated by its photochemical product SO$_2$. We note that the non-detection of H$_2$S for this specific case, which stands in contrast to the detection of H$_2$S in the main text, is a result of the higher amounts of CH$_4$ assumed in the simulated case, which masks the H$_2$S. As such the first part of the cascade is completed, with the information propagating to subsequent stages being the canonical set of retrieved species.

In addition to the mixing ratios of the above chemical species, the retrieval also constrains the properties of the parametric grey clouds and Rayleigh-like hazes. It specifically constrains the Rayleigh enhancement factor and scattering slope to very large values, at log($a$)$ =9.66^{+0.25}_{-0.48}$ and $\gamma = -13.04^{+0.70}_{-0.45}$. This is either because of the presence of a very strong scattering slope in the optical, or hazes are being invoked to explain the non-grey nature of aerosol contributions at longer wavelengths.

\subsection{Refining The Aerosol Model}

As a result of the first step of the cascade, our atmospheric model is sufficient to fit all prominent absorption features. This allows us to now consider fitting the spectral baseline better by refining the atmospheric aerosol model. Such refinement is warranted, as the prior retrievals have indicated that there is significant spectral contribution from atmospheric aerosols, which may not be well characterised by grey clouds.

\begin{figure}
    \centering
    \includegraphics[angle=0,width=0.95\linewidth]{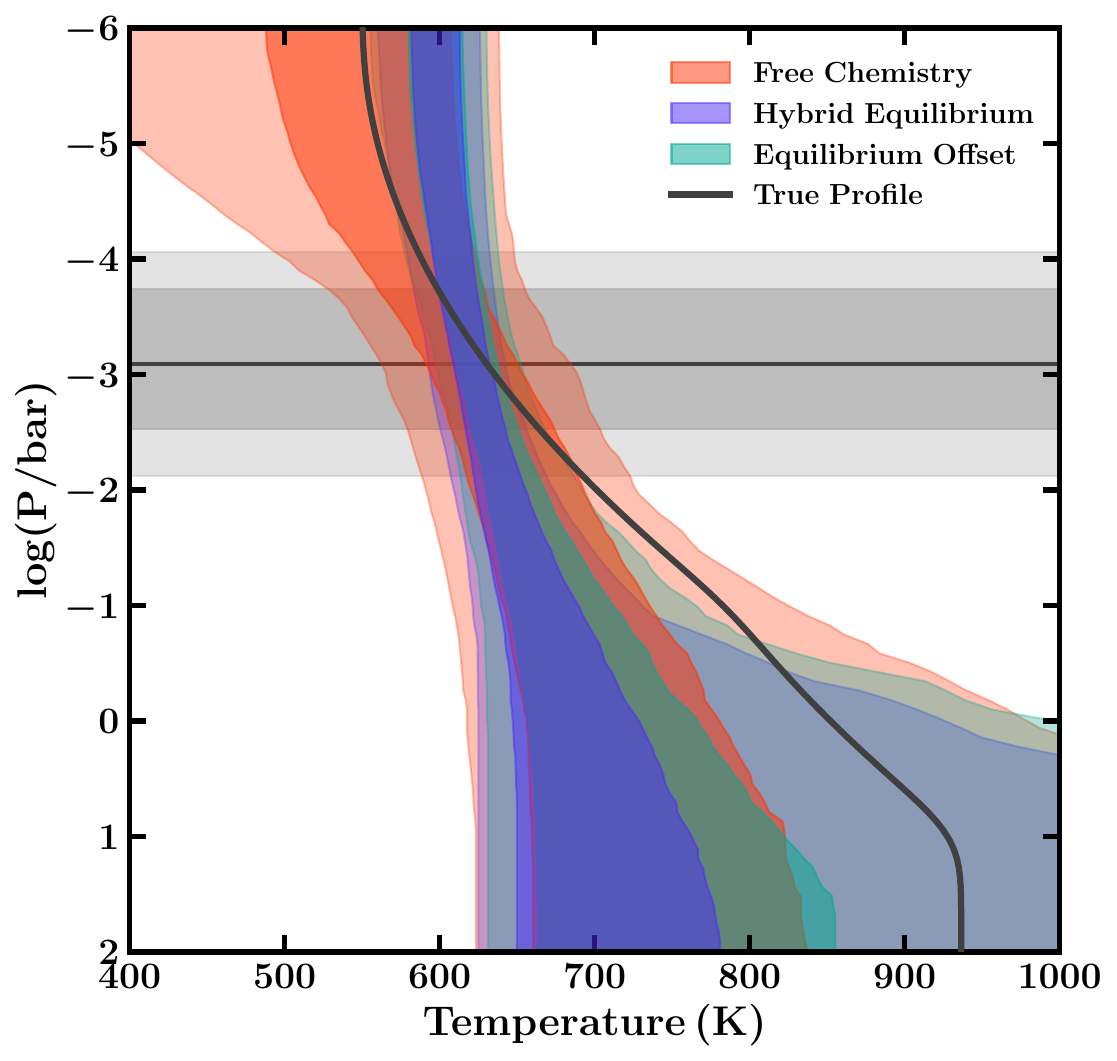}
    \caption{ The P-T profile used to generated the simulated observations (grey line) and the spectral fits obtained with the three atmospheric chemistry parametrisations, where darker and lighter shaded regions show the 1- and 2-$\sigma$ contours. The horizontal black line and grey shaded areas show the median, 1- and 2-$\sigma$ photospheric pressures across the observed wavelength range.}
    \label{vira_fig:simulated_PT_comparison}
\end{figure}

\subsubsection{Sigmoid Clouds}

Our first step is to establish and quantify the non-grey nature of a possible cloud deck. We therefore use the sigmoid clouds model, which also retains the Rayleigh-like haze component. As such, this can indicate if the prior constraints are the result of a strong scattering slope or non-grey clouds.

This retrieval obtains precise constraints for the sigmoid mid-point of 1.03$^{+0.05}_{-0.05}$ $\mu$m and decay width of $6.94^{+1.22}_{-1.00}$~$(\mu\mathrm{m})^{-1}$, indicating that there is indeed spectral contribution from non-grey clouds. Moreover, this retrieval does not constrain the Rayleigh-like haze parameters, indicating that non-grey clouds are a more favourable explanation of the data rather than a very strong scattering slope.

\begin{figure*}
    \centering
    \includegraphics[angle=0,width=\textwidth]{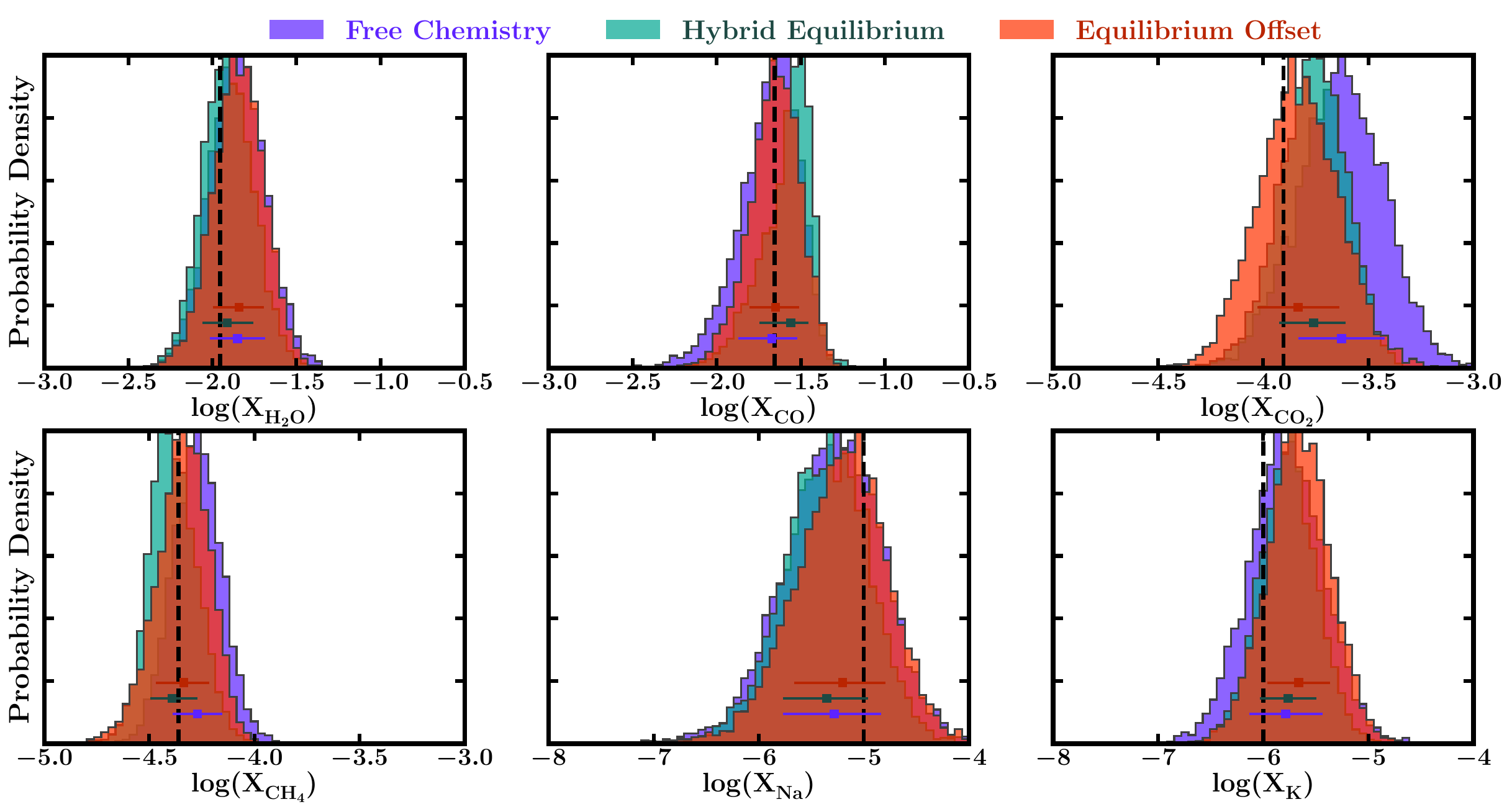}
    \caption{Posterior probability distributions for the retrieved mixing ratios of H$_2$O, CO, CO$_2$, CH$_4$, Na and K, averaged over the observed photosphere between 10$^{-2}$-10$^{-4}$~bar. For each species, three posterior distributions are shown, corresponding to the free chemistry, hybrid equilibrium and equilibrium offset stages of teh cascade, all with Mie scattering aerosols.}
    \label{vira_fig:appendix_simulated_posteriors}
\end{figure*}

\subsubsection{A Canonical Set of Aerosol Species}

With the knowledge that there is significant non-grey cloud absorption affecting the spectral baseline, we can now consider more realistic Mie scattering cloud models. We now consider which aerosol species to include in our model, in a similar manner to how the set of included gaseous species was determined. With a maximal set of aerosol species, the retrieval obtains constraints for the mixing ratio of ZnS aerosols. At the same time, the mixing ratio posterior distribution for MgSiO$_3$ is reasonably constrained, but with a very significant tail towards lower abundances. As such, we include both in the retrieval model. Additionally, we include KCl, as its condensation profile is very similar to that of ZnS \citep{Morley2013} and could theoretically be expected to also be present alongside ZnS.

With the canonical set of aerosol species, the retrieval obtains constraints that are largely consistent with the true values used to generate the observations: $X_{\mathrm{ZnS}} = 7.55^{+0.76}_{-0.93}$, $\mathrm{log}(r_\mathrm{c} / \mu \mathrm{m}) = -2.40^{+0.28}_{-0.23}$, $H_\mathrm{c} = 0.94^{+0.05}_{-0.07}$ and $f_\mathrm{c} = 0.50^{+0.04}_{-0.03}$. The retrieved P-T profile, shown in Figure \ref{vira_fig:simulated_PT_comparison}, is consistent with the true P-T profile that was used to generate the data to within 1-$\sigma$ across the observed photosphere. 

\subsection{Hybrid Equilibrium}

Thanks to the preceding stages of the cascade, the atmospheric model is now capable of explaining all absorption peaks while also obtaining a good fit to the spectral baseline. Motivated by the high signal-to-noise observations, we now explore refinements to the atmospheric composition model which consider non-uniform abundance profiles.

We begin with a hybrid equilibrium retrieval, in order to obtain a first estimate of the elemental abundances which can be passed to subsequent stages. As in the main body of this work, we set the vertical mixing ratios of H$_2$O, CH$_4$, CO, CO$_2$, HCN, C$_2$H$_2$ and H$_2$S to be determined by the equilibrium chemistry calculation, while those of SO$_2$, Na and K are left as free-chemistry vertically constant profiles. 

The retrieval constrains the abundances of $[\frac{\mathrm{O}}{\mathrm{H}}]$ and $[\frac{\mathrm{C}}{\mathrm{H}}]$ to $+1.23^{+0.14}_{-0.13}$ and $+1.05^{+0.08}_{-0.07}$, respectively. We therefore note that $[\frac{\mathrm{C}}{\mathrm{H}}]$ is consistent with the true value to 1-$\sigma$, $[\frac{\mathrm{O}}{\mathrm{H}}]$ is not. Meanwhile the retrieved P-T profile, shown in Figure \ref{vira_fig:simulated_PT_comparison}, is largely consistent with the true profile in the photosphere. Compared to that obtained with the free chemistry retrieval, the retrieved P-T profile is more precise but shows a less steep temperature increase, due to the coupling between the temperature profile and atmospheric abundances.

It is notable that this retrieval over-estimates $[\frac{\mathrm{O}}{\mathrm{H}}]$ rather than $[\frac{\mathrm{C}}{\mathrm{H}}]$, in spite of the fact that carbon-bearing molecules CH$_4$ and CO are the ones that had been depleted and enhanced, respectively. As such,

\subsection{Equilibrium Offset}

The first-guess elemental abundance constraints for $[\frac{\mathrm{O}}{\mathrm{H}}]$ and $[\frac{\mathrm{C}}{\mathrm{H}}]$ are now used in the equilibrium offset retrievals that comprise the final steps of the cascade. We specifically fix $[\frac{\mathrm{O}}{\mathrm{H}}]$ +1.2 and $[\frac{\mathrm{C}}{\mathrm{H}}]$ to +1.0. While unconstrained in the hybrid equilibrium retrieval, we also nominally fix $[\frac{\mathrm{N}}{\mathrm{H}}]$ and $[\frac{\mathrm{S}}{\mathrm{H}}]$ to +1.

The starting point for the equilibrium offset parameters is different to those that generated the simulated observations, which therefore precludes a direct comparison of the retrieved offset values to those used to generate the simulated spectrum. At first glance, however, the retrieval correctly finds that CO is enhanced by $\sim$1 dex above equilibrium values, while CH$_4$ is depleted by $\sim$2~dex. Additionally, it also constrains the H$_2$O abundance offset to values corresponding to a slight depletion, which is a consequence of the provided initial $[\frac{\mathrm{O}}{\mathrm{H}}]$ guess being higher than the true value. The retrieved P-T profile is again consistent with the true profile in the photosphere, as seen in figure \ref{vira_fig:simulated_PT_comparison}. It is constrained with a precision between that of the hybid equilibrium and free chemistry retrievals, as the coupling between the atmospheric chemistry and temperature structure is present but not as strong as in the hybrid equilibrium case.

We further assess the performance of the retrieval by considering the averaged mixing ratio of each species across the effective photosphere, which in this case lies between 10$^{-2}$-10$^{-4}$~bar, and comparing that to the similarly averaged true values. This is shown in Figure \ref{vira_fig:appendix_simulated_posteriors}. It can be seen that the retrieval successfully compensated for the somewhat inaccurate $[\frac{\mathrm{O}}{\mathrm{H}}]$ and $[\frac{\mathrm{C}}{\mathrm{H}}]$ first guesses provided, obtaining atmospheric abundance constraints that are consistent with the simulated atmosphere to within 1-$\sigma$. Our findings therefore demonstrate that the equilibrium offset approach is robust against minor inaccuracies in the fiducial elemental abundances provided from prior stages of the cascade.

We finally confirm the findings of the equilibrium offset retrieval with the final stage of the cascade, where we replace the MultiNest Nested Sampling implementation with UltraNest to confirm the accuracy of our constraints. This final retrieval yields abundance constraints that are highly consistent with those obtained with MultiNest. As such, we have reached the end of the retrieval cascade.

As can be seen in Figure \ref{vira_fig:appendix_simulated_posteriors}, going from free chemistry to hybrid equilibrium and finally equilibrium offset affected the exact constraints obtained. While in most cases all three composition parametrisations give rise to constraints that are consistent with the true values to within 1-$\sigma$, those of CO$_2$ show greater scatter with only the final equilibrium offset retrieval producing an accurate constraint.


\bsp	
\label{lastpage}
\end{document}